\documentclass[sigconf]{acmart}

\usepackage{verbatim}

\AtBeginDocument{%
  }

\usepackage[leftmargin=0.5cm, rightmargin=0.5cm]{quoting}
\usepackage{enumitem}
\usepackage{afterpage}
\usepackage{graphicx}
\usepackage{caption}
\usepackage{subcaption}

\newcounter{fcounter}
\newenvironment{italicquote}
    {\noindent\begin{quoting}\itshape} 
    {\end{quoting}} 

\newcommand{\italicquoteinline}[1]{``\textit{#1}''}

\usepackage{ifthen}
\newboolean{showedits}
\setboolean{showedits}{false} 

\ifthenelse{\boolean{showedits}}
{
	\newcommand{\del}[1]{\textcolor{red}{\sout{#1}}} 
}{
	\newcommand{\del}[1]{} 
	
}

\newboolean{showcomments}
\setboolean{showcomments}{false} 
\newcommand{\id}[1]{$-$Id: scgPaper.tex 32478 2010-04-29 09:11:32Z oscar $-$}

\ifthenelse{\boolean{showcomments}}
{\newcommand{\nbc}[3]{
 {\colorbox{#3}{\bfseries\sffamily\scriptsize\textcolor{white}{#1}}}
 {\textcolor{#3}{\sf\small$\blacktriangleright$\textit{#2}$\blacktriangleleft$}}}
 }
{\newcommand{\nbc}[3]{}
 \renewcommand{\del}[1]{} 
 }
 \definecolor{darkyellow}{RGB}{255, 222, 9}
 
\copyrightyear{2026}
\acmYear{2026}
\setcopyright{cc}
\setcctype{by-nc-nd}
\acmConference[CHI '26]{Proceedings of the 2026 CHI Conference on Human Factors in Computing Systems}{April 13--17, 2026}{Barcelona, Spain}
\acmBooktitle{Proceedings of the 2026 CHI Conference on Human Factors in Computing Systems (CHI '26), April 13--17, 2026, Barcelona, Spain}
\acmDOI{10.1145/3772318.3791310}
\acmISBN{979-8-4007-2278-3/2026/04}

\usepackage{tabularx} 
\usepackage{array}    
\usepackage{booktabs} 
\usepackage{enumitem} 

\newcolumntype{Y}{>{\raggedright\arraybackslash}X}

\usepackage{xcolor}
\usepackage{hyperref}

\hypersetup{
    colorlinks,
    linkcolor=blue,     
    filecolor=magenta,  
    urlcolor=cyan,      
    citecolor=green,    
}

\usepackage{acmart-taps}

\makeatletter
\def\@authorfont{\Large\sffamily}
\def\@affiliationfont{\small\normalfont}
\makeatother

\begin{document}

\title{Crystallizing Schemas with \textit{Teleoscope}: Thematic Curation of Large Text Corpora on Reddit}

\author{Patrick Yung Kang Lee}
\authornote{The first two authors contributed equally to this work.}
\orcid{0000-0002-3385-5756}
\affiliation{
\department{Computer Science}\institution{University of Toronto}
\city{Toronto}
\state{Ontario}
\country{Canada}}
\email{patricklee@cs.toronto.edu}

\author{Paul Hendrik Bucci}
\authornotemark[1]
\orcid{0000-0002-8646-7730}
\affiliation{
\department{Computer Science}\institution{University of British Columbia}
\city{Vancouver}
\state{British Columbia}
\country{Canada}}
\email{pbucci@cs.ubc.ca}

\author{Leo Itsuki Foord-Kelcey}
\orcid{0009-0007-5418-6778}
\affiliation{
\department{Computer Science}\institution{University of British Columbia}
\city{Vancouver}
\state{British Columbia}
\country{Canada}}
\email{leofk@cs.ubc.ca}

\author{Alamjeet Singh}
\orcid{0009-0005-6650-3377}
\affiliation{
\department{Computer Science}
\institution{University of British Columbia}
\city{Vancouver}
\state{British Columbia}
\country{Canada}}
\email{singhalamjeet18@gmail.com}

\author{Ivan Beschastnikh}
\orcid{0000-0003-1676-8834}
\affiliation{
\department{Computer Science}\institution{University of British Columbia}
\city{Vancouver}
\state{British Columbia}
\country{Canada}}
\email{bestchai@cs.ubc.ca}

\begin{abstract}
Large text corpora, such as Reddit posts, have become an increasingly prevalent site of qualitative inquiry. However, most large text corpora are intractable for qualitative researchers. Instead, teams rely on statistical subsampling to reduce corpora to a manageable size for qualitative analysis. While previous work for navigating large corpora involves visualizing the dataset at the corpus-level using high-level statistical summaries, few systems offer the ability to curate data using an interpretivist approach. To address this, we developed \textit{Teleoscope}, a web-based interface designed to scaffold iterative, interactive, and reflexive refinement of a large corpus, in a process we call \textit{thematic curation}. Across three deployments, we learned that \textit{Teleoscope} supports serendipitous discovery of new keywords, results in greater feelings of confidence in search saturation, and aids collaborative discussion of alternative curation pathways. \textit{Teleoscope} empowers researchers to stay "close to the data" in order to make qualitative workflows methodologically coherent with large text corpora.
\enlargethispage*{16pt}\end{abstract}


\begin{teaserfigure}
\centering
\includegraphics[width=0.81\textwidth]{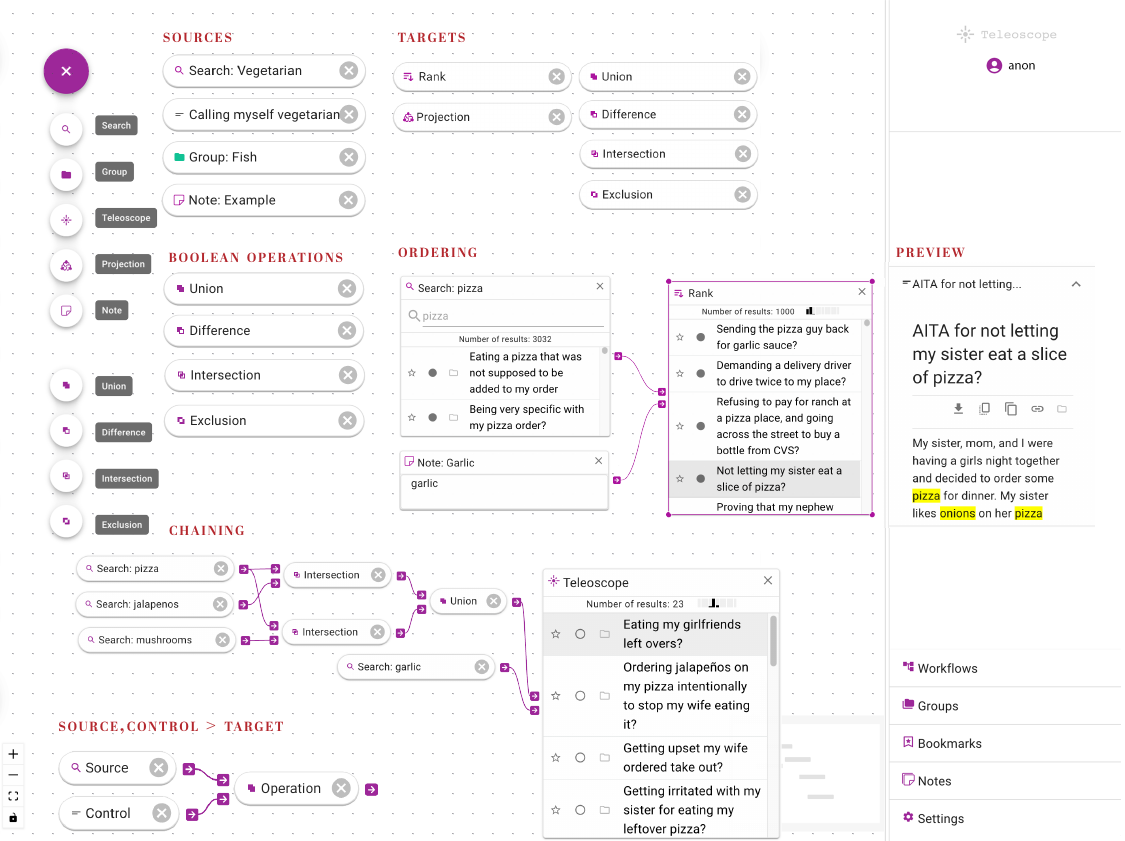}
    \caption{\textit{Teleoscope} is a web-based interface for performing \textit{thematic curation} of large text corpora, such as social media data from Reddit. By iteratively refining user-defined metrics of similarity, \textit{Teleoscope} supports an interpretivist approach to data curation. This helps address the problem of ad-hoc workflows that rely on statistical sampling, which are epistemically misaligned with interpretivist approaches to qualitative analysis.}
    \label{fig:teaser}
    \Description{Teaser figure showing the \textit{Teleoscope} interface. \textit{Teleoscope} is a web-based interface for performing \textit{thematic curation} of large text corpora, such as social media data from Reddit. By iteratively refining user-defined metrics of similarity, \textit{Teleoscope} supports an interpretivist approach to data curation. This helps address the problem of ad-hoc workflows that rely on statistical sampling, which are epistemically misaligned with interpretivist approaches to qualitative analysis.}
\end{teaserfigure}


\maketitle

\section{Introduction}
\label{sec:Introduction}
Exploring and analyzing large text corpora has become an increasingly prevalent mode of qualitative inquiry for researchers in HCI \cite{gauthier2022will}, healthcare \cite{low2020natural, slemon2021reddit}, and the social sciences \cite{maxwell2020short}. Qualitative research often focuses on contextual interpretation, reflexivity, and researcher positionality  ~\cite{hennink2017code, sebele2020saturation, ladonna2021beyond}. The primary goal of this approach is to provide rich, nuanced insights into social or cultural phenomena that more quantitative/positivist approaches fail to capture. A popular data source for qualitative data is the social media platform Reddit, since it contains many normative and descriptive examples of social interaction and cultural norms. However, the corpora present on Reddit are usually too large to be tractable for small qualitative research teams. Instead, researchers are forced to rely on ad-hoc or statistical sub-sampling strategies. This causes researchers to miss a powerful opportunity to qualitatively curate data based on an interpretive approach that better aligns with the core tenets of qualitative research .

In recent years, advances in AI and ML have opened up new, albeit somewhat controversial, opportunities for qualitative analysis \cite{jiang_supporting_serendipity, schroeder_llms}. In particular, modeling semantics using contextual dependencies in language data is now being realized at scale with large language models (LLMs), providing an avenue for qualitatively curating data. Although prior work in semantic search interfaces has excelled at providing topic-based overviews of the entire dataset \cite{Hong2022}, inductive approaches to document curation (i.e., starting from the document-level rather than the dataset-level for curation) remain relatively unexplored. In this paper, we introduce the idea of \textit{thematic curation}, an inductive, interpretivist approach to corpus construction. Using vector embeddings~\cite{bge_m3} between related documents, researchers can iteratively curate a large text corpus into a tractable collection of documents for downstream qualitative analysis. We call this collection, supported by a traceable network of document interconnections, a \textbf{thematic schema}, which corresponds to a researcher's mental model (cognitive schema) of the relationships and meanings between documents. 

To enable thematic curation, we built \textit{Teleoscope}, a web-based system designed to support multiple users interactively exploring and refining a large corpus of Reddit documents (100K-1M posts). Before conducting analysis proper \cite{braun2006using, braun2012thematic, muller2014curiosity}, qualitative researchers working with large text corpora typically undertake data curation \cite{karcher2021data, mannheimer2021data}, and sense-making \cite{berret2022iceberg}. Data curation involves the task of sampling that happens before any qualitative study of text. Sense-making refers to the process of an analyst "making sense" of the data in their own minds, which often happens in parallel with curation and analysis. \textit{Teleoscope} assists qualitative researchers during the data curation and sense-making phases of a qualitative research study. \textit{Teleoscope} functions by allowing researchers to browse and navigate their corpus on an infinite canvas via an initial keyword search and then refining opinionated, interpretive document groupings. When using \textit{Teleoscope}, a user's curatorial process results in visual artifacts that trace their thought process in the form of action histories and the curated schemas, allowing researchers to compare their curation process with that of other research team members. The visual traces enable discussions of provenance, that is, how a researcher curated their dataset according to their particular positionality, helping to bring differences between team member's interpretations to the surface for discussion before analysis begins.

The design process for \textit{Teleoscope} spanned several years, including a series of prototypes, formative studies, and focus groups that informed the final system design. We conducted an extended formative study with a "simulated" research team (n=5), asking qualitative researchers to use \textit{Teleoscope} to explore a shared topic over the course of six weeks. The simulated research team was then convened to discuss their experience and findings at a focus group.
We then recruited a three-person qualitative research team of nurses for a field deployment where they used \textit{Teleoscope} as part of their data curation and analysis process and responded to their design requests. We have also released a public, cloud-native version of \textit{Teleoscope} \cite{teleoscope2024postreview, teleoscope2024githubpostreview} that has had users in market research and public policy who have provided feature requests.

We learned through our deployments that \textit{Teleoscope} was able to support serendipitous discovery of previously unknown keywords and terms for phenomena of interest among qualitative researchers. It also resulted in greater feelings of confidence and rigour in search saturation among participants while assuaging feelings of interpretive bias. In addition, being able to collaboratively explore alternative pathways to data curation opened up nuanced discussions about the role different interpretive lenses could have on the final curated dataset. We discuss the implications of these findings on working with large text corpora in interpretivist paradigms of qualitative research. We also discuss learnings from observing differences between top-down and bottom-up approaches to data exploration more broadly.

Our contributions in this paper include:
\begin{itemize}[leftmargin=*,label=$\star$]
    \item \textbf{C1. Methodological analysis of a highly interpretive sampling strategy we call \textit{thematic curation}, resulting in outputs we call \textit{thematic schemas}.}
    \item \textbf{C2. An interpretivist data curation platform that is open-source and supports collaboration.}
    \item \textbf{C3. Empirical results of deployments with qualitative researchers working with large text corpora.}
\end{itemize}

\section{Related Work}
\label{sec:Related Work}
We motivate the need for a system like \textit{Teleoscope} by discussing work related to the growing use of qualitative analysis techniques on large datasets, the role of computational methods in supporting qualitative analysis, and the way \textit{Teleoscope} builds on work in interactive data exploration systems.

\subsection{Qualitative Analysis on Large Datasets}
\label{sec:RW Qual Analysis}
Large text corpora, such as those created by social media platforms, have become an increasingly prevalent site for qualitative research across disciplines \cite{maxwell2020short, low2020natural, slemon2021reddit, gauthier2022will}. In particular, during the COVID-19 pandemic, when face-to-face research was not possible due to quarantine and stay-at-home measures, sites like Reddit became an important source for qualitative academic work \cite{norman2024scraping}. However, analyzing large text corpora, such as those found on Reddit, presents significant scalability challenges for small qualitative research teams. Even a modestly-sized corpus, between 10K to 100K documents, is often intractable to a research team if they wish to exhaustively examine its contents. As a result, some degree of data curation is required to make the dataset amenable to qualitative analysis.

In practice however, the choice of data curation strategy is often highly specific to the research project, and ad-hoc sub-sampling or partitioning strategies are often employed due to a lack of agreed upon standards for curation. For example, Low et al. \cite{low2020natural} use a random sub-setting strategy on their datasets to significantly reduce the number of posts that need to be analyzed by their topic modeling system. More principled or semi-structured down-selection strategies come with their own limitations --- whether because they presuppose certain aspects of the document space, such as stratified sampling's \cite{singh1996stratified} assumption of the existence of discrete, mutually exclusive strata, or the active sampling \cite{laws2011active} assumption of a learnable decision boundary; or provide limited interpretive flexibility for human curatorial decisions, such as in diversity-based sampling \cite{de2011find} which prioritizes coverage of the document space rather than intentional curatorial decisions, or topic-guided sampling \cite{bakharia2016interactive}, which allows only for interpretation at the algorithmically-provided topic level. These strategies introduce tensions between the way the data is curated --- often via a statistical, positivist approach --- and the way the data is later analyzed --- via a qualitative, interpretivist approach. Positivist analysis aims to evaluate specific and objective hypotheses via observation and measurement, differing significantly from qualitative analysis in terms of how universality, generalizability and rigour are conceived \cite{soden2024evaluating}. Qualitative work, especially those drawing from an interpretivist paradigm, aim to present "rich descriptions of participant experiences, along with the context specificity and depth associated with how participants interpret and understand those experiences" \cite{soden2024evaluating}. Providing the opportunity to scaffold iterative, interactive, and reflexive sampling that prioritizes the interpretive skills of the researcher while providing them with computational and statistical tooling that allows them to meaningfully explore an otherwise intractable corpus is a gap in current qualitative approaches.



\begin{figure*}[p]
  \centering
  \includegraphics[width=0.85\textwidth]{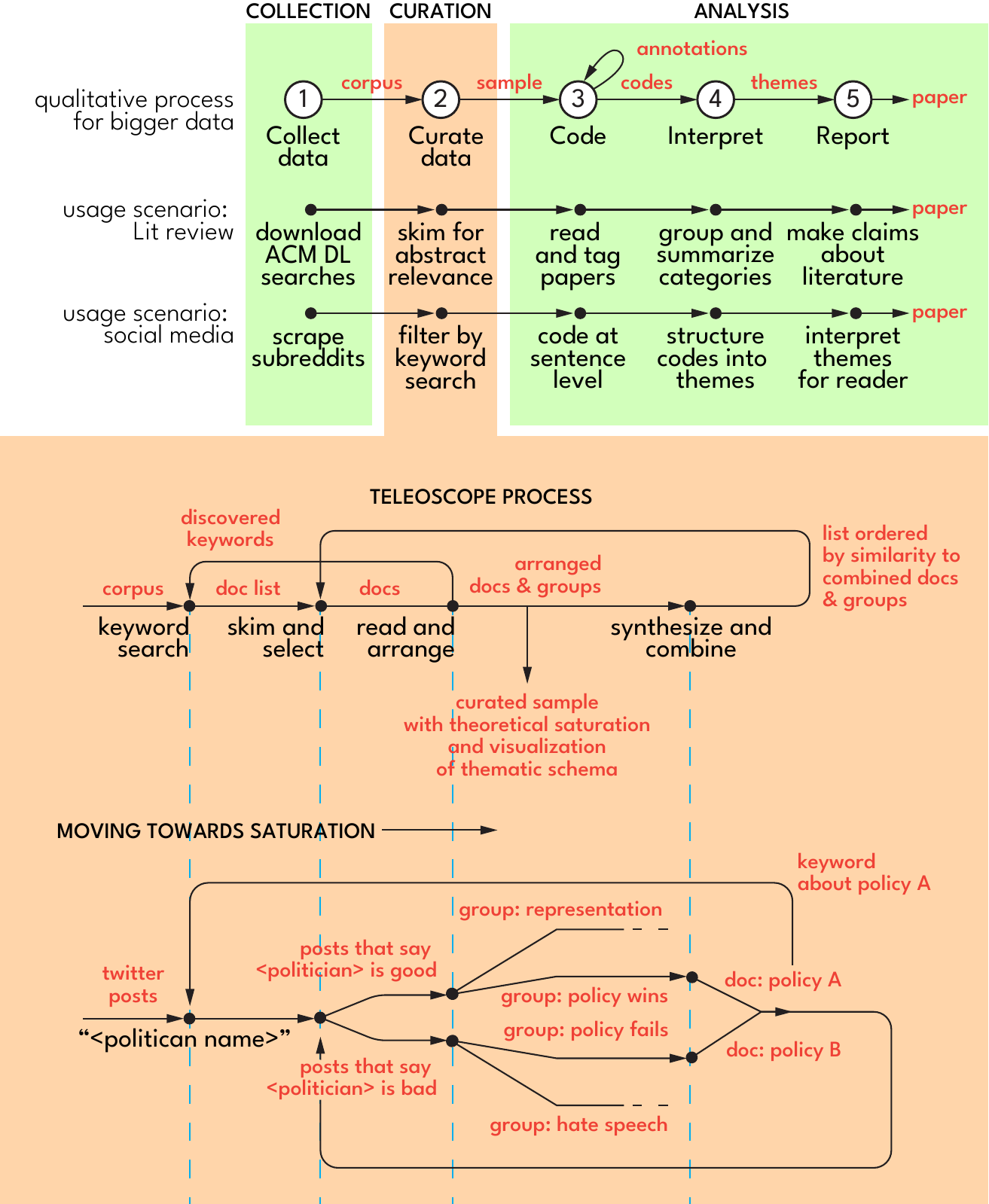}
  \caption{The \textit{Teleoscope} process is part of the data curation phase of qualitative research. Users start by a keyword search to filter the corpus and initially discover documents, which they skim and select the best ones to put on the workspace. They then read more closely and rearrange the documents on the workspace, eventually arranging some into groups, while leaving others nearby for re-discovery. Synthesizing their understanding of the documents, they choose documents or groups to combine into a similarity search, starting the process of skimming again. When a new keyword is discovered, the corpus can be filtered again. Saturation is achieved when no new keywords can be found; signpost documents show differences between groups are identified, and groups are thematically homogeneous.}
  \Description{Diagram displaying the Teleoscope process. The top half of the diagram displays the collection, curation, and analysis steps of the process. The curation portion of the process is enlarged to show the steps unique to the Teleoscope process. These include keyword searches, skimming capabilities, being able to arranged documents, and allowing for synthesis and combination.}
  \label{fig:teleoscope-process}
\end{figure*}

\subsection{AI in Qualitative Analysis}
\label{sec:RW AI}
Due to rapid advances in NLP, machine learning, and LLM-powered tooling, developing AI-powered qualitative analysis tools has rapidly become an active area of research. Although many of these systems predate the advent of LLMs \cite{yan2014optimizing, gebreegziabher2023patat, gao2023coaicoder, overney2024sensemate}, the ability of LLMs to model semantic context at-scale has spurred the development of many others \cite{suh2023sensecape, gao2024collabcoder, sharma2025details}. The systems that are most similar to \textit{Teleoscope} are Scholastic \cite{Hong2022} and NEEDLE \cite{NEEDLE}, which provide novel ways for researchers to remain "close" to their datasets in their workflows, by drilling down to individual documents or supporting claim-based document retrieval, respectively. \textit{Teleoscope} builds on these systems by providing a different orientation towards data, using machine learning workflows to iteratively curate the document space into \textbf{thematic schemas} (defined below) using user-selected examples, rather than through random sub-setting or by imposing a relational claims-based exploration of the corpus. A system that shares similar design goals is the Apolo system \cite{chau2011apolo}. Like \textit{Teleoscope}, Apolo provides an interactive visualization system for sense-making with large network datasets. However, \textit{Teleoscope's} target datasets (unstructured social media data) are quite different from the networked datasets Apolo was designed to explore, though they share some interaction commonalities, particularly the use of ML techniques to provide in-place ranking of documents of interest to the user. In addition, \textit{Teleoscope} builds upon a screen space limitation in the original Apolo interface by using an infinite canvas to help manage screen real estate, whereas the Apolo system could only handle two themes in their user study. While performing baseline comparisons, we found that, although many commercial qualitative research tools like NVivo now include AI support, most focus on coding rather than curation (a full comparison is available in the supplementary materials).

We draw on literature from psychology on cognitive schemas to define our concept of \textbf{thematic schemas}. In psychology, a cognitive schema is defined as a set of abstract beliefs that a person develops in response to life experiences \cite{piaget2005language, berret2022iceberg}. We extend the idea of cognitive schemas to beliefs about data, resulting in \textbf{thematic schemas}: Curated collections of documents supported by a traceable network of document interconnections that correspond to cognitive schemas/mental models a researcher has about the relationships and meanings between documents. \textit{Thematic schemas} can then be used as the data source for a full-scale qualitative analysis downstream. To be clear, thematic schemas are not complete collections of evidentiary data for a theme of the sort that is conceived as the end result of processes like reflexive thematic analysis \cite{braun2006using}. Rather, they are the output of employing a principled and systematic data exploration and curation process on a much larger dataset that reify internal cognitive schemas/mental models a researcher possesses.

To facilitate the creation of thematic schemas, which possess greater interpretive depth and semantic richness than simple topic categories, \textit{Teleoscope} takes a modeling by example approach \cite{lissandrini2019data} to allow users to explore and structure their corpus. These systems \cite{martins2019reverse, fariha2019example, lissandrini2019data} allow queries to be synthesized from examples. In a similar fashion, \textit{Teleoscope} uses semantic vector similarity to drive exploration. However, \textit{Teleoscope} differs from these systems by avoiding the construction of deterministic queries or by modeling the entirety of the corpus. Instead, it relies on the interaction process and the researcher's own sense of information saturation (see Figure~\ref{fig:teleoscope-process}) to determine the extent of exploration, in line with interpretivist values in qualitative research \cite{braun2019reflecting}.


In addition to the data curation step, LLMs are increasingly being used throughout the qualitative research life-cycle, from research ideation, to the drafting of research materials, to qualitative coding \cite{schroeder_llms}. \textit{Teleoscope} is meant to assist with the specific task of sub-setting large datasets, such as those in public social media forums into tractable datasets that can then have more traditional qualitative analysis methods applied to them. When a user has finished a phase of exploration using semantic similarity (i.e., via Search and Rank operations, see Section~\ref{sec:Features}), they can switch to structuring the document space via semi-supervised dimensionality reduction. Dimensionality reduction has been a popular technique in many human-centered ML and visualization systems \cite{el2019semantic, meinecke2021explaining, asudani2023impact, sperrle2021survey} owing to its ability to make large datasets more tractable to human manipulation. \textit{Teleoscope} leverages these techniques to provide a system for interactive and serendipitous data curation that fosters researcher agency \cite{jiang_supporting_serendipity}. The diagram in Figure~\ref{fig:teleoscope-process} illustrates the phase in qualitative research workflows \textit{Teleoscope} is meant to support, relative to other common phases, such as data collection, coding, and analysis.\enlargethispage*{16pt}

\subsection{Interactive Data Exploration Systems \& Provenance}
\label{sec:RW Viz}
The HCI community has developed a large body of work in interactive systems for data visualization and exploration \cite{kucher2015text, hearst2009search, jones1999graphical, cutting2017scatter}. However, in the context of qualitative analysis, existing systems are either primarily designed to assist with the analysis phase, or are incapable of scalable search to large corpora. \textit{Teleoscope} focuses on capturing and visualizing the exploratory \textit{process} ~\cite{hearst2013sewing, li2023thinking} involved in data curation, an area referred to as \textit{provenance} tracking in systems and visualization literature ~\cite{xu2020survey, yuan2021survey}. \textit{Teleoscope's} focus on visualizing provenance stems from a desire to make data curation reproducible and explainable ~\cite{yang2019unremarkable, silva2007provenance, arrieta2020explainable} to help with the process of analysis, publication, and review. In contrast, many topic model visualizations choose instead to directly visualize the underlying document clusters, the \textit{result} of clustering~\cite{nikolenko2017topic, rudiger2022topic}.

To make sense of qualitative data, it is important to not only answer the question of \emph{``What are the results?''} but also \emph{``How did we get here?''}. This can be achieved by creating a visual trace or history of user interactions with the data.   In \textit{Teleoscope}, keeping track of the provenance of a thematic schema involves tracking the inputs and user-selected data processing elements in a chained workflow. The user is engaged in a process of data wrangling for sense-making~\cite{bors2019capturing, badi2006recognizing}, i.e., \textit{exploration}, \textit{annotation} and \textit{curation} of data~\cite{munzner2014visualization}. In \textit{Trrack}~\cite{cutler2020}, Cutler et al. demonstrate a library for tracking branching histories for actions on web-based visualizations \cite{yuan2021survey}. This is a maximal approach for provenance, where every action is tracked and is reproducible.
 
Although \textit{Teleoscope} logs every user action in its history system, most of this is hidden from the user. Instead, we present a sparse, non-linear canvas interface where individual documents, groupings, and schemas can be directly manipulated by the user to foster a sense of spontaneous and serendipitous sense-making in the data curation process for qualitative researchers. This allows users to create iterative, comprehensible workflows that they commit to by grouping and connecting elements in a workflow graph. In Xu et al.'s taxonomy of topic model approaches \cite{xu2020survey}, we are modeling coupled user and application state via an entity graph using semantic interactive visual analysis. This approach also has historical roots in search and information seeking interfaces, such as the Yahoo Pipes system \cite{pruett2007yahoo} and the work of Endert et al. on user-generated spatializations of text documents \cite{endert2012semantics}.

The rest of the paper is organized as follows: Section 3 describes our use of Research through Design (RtD) principles \cite{zimmerman2007research} to conduct a series of formative studies and design prototypes that led to our design goals for \textit{Teleoscope}. Section 4 then presents the system design and usage metaphors of \textit{Teleoscope} in full. Section 5 describes the results of evaluating \textit{Teleoscope} with a second extended formative study over the course of several weeks and a field deployment of \textit{Teleoscope} with a research team that used \textit{Teleoscope} over several months. All user studies conducted over the course of developing \textit{Teleoscope} are summarized in Table~\ref{tab:study_summary}. Section 6 discusses the role thematic curation has to play in interpretivist paradigms of qualitative research that seek to use large text corpora as their main data source and the affordances of top-down versus bottom-up approaches to corpus visualization.\enlargethispage*{16pt}

\section{Interface Design Process}
\label{sec:Formative Study}
We took inspiration from Research through Design (RtD) principles \cite{zimmerman2007research} to iteratively explore the ideas that resulted in our eventual design goals and interface for a thematic curation system. In the first year of developing this project, \textit{Teleoscope} went through a series of prototypes and design iterations. During this time, we experimented with features such as summary statistics, different NLP approaches, and dashboard-style interfaces, but quickly found these were unsuitable for interpretivist approaches to data curation that qualitative researchers we talked to wanted. This was due to a desire among qualitative researchers to remain "close to the data" \cite{byrne2022worked}, rather than reading quantitative metrics or wholesale summaries that deprive them of the opportunity to generate interpretive insights based on specific documents in the corpus.

\begin{table*}[ht]
\centering
\begin{tabularx}{\textwidth}{lYYYYY}
\toprule
\textbf{Study Type} & \textbf{Participants} & \textbf{Dataset/Topic} & \textbf{Methods} & \textbf{Takeaways} \\
\midrule
Formative Study &
\begin{itemize}[leftmargin=*,nosep]
  \item Computer scientists (undergraduate to post-doc)
\end{itemize} &
\begin{itemize}[leftmargin=*,nosep]
  \item Privacy and security norms on r/AITA ($\sim$350K documents)
\end{itemize} &
\begin{itemize}[leftmargin=*,nosep]
  \item Iterative prototyping
  \item Low-cost UX evaluation
  \item Wizard-of-Oz testing
\end{itemize} &
\begin{itemize}[leftmargin=*,nosep]
  \item Researchers want to work "close to the data"
  \item Need support for organizing corpora interpretively
\end{itemize} \\
\addlinespace
Extended Formative Study &
\begin{itemize}[leftmargin=*,nosep]
  \item PI in Nursing
  \item PhD student in sociology
  \item Three RAs
\end{itemize} &
\begin{itemize}[leftmargin=*,nosep]
  \item Critical and end of life care on r/AITA ($\sim$350K documents)
\end{itemize} &
\begin{itemize}[leftmargin=*,nosep]
  \item Briefing session
  \item Diary study (10 hours of use)
  \item Post-study interviews
\end{itemize} &
\begin{itemize}[leftmargin=*,nosep]
  \item Quick, iterative curation shows promise
  \item Better support needed for nuanced interpretations
\end{itemize} \\
\addlinespace
Field Deployment &
\begin{itemize}[leftmargin=*,nosep]
  \item PI in nursing
  \item Two graduate RAs from schools of nursing
\end{itemize} &
\begin{itemize}[leftmargin=*,nosep]
  \item Structural inequality among nurses on r/nursing ($\sim$35K documents)
\end{itemize} &
\begin{itemize}[leftmargin=*,nosep]
  \item Briefing session
  \item In-the-wild study
  \item Semi-structured interviews
\end{itemize} &
\begin{itemize}[leftmargin=*,nosep]
  \item Helps discover terminology
  \item Fosters discussion on alternative curation pathways
  \item Speeds up saturation on large datasets
\end{itemize} \\
\addlinespace
Public Deployment &
\begin{itemize}[leftmargin=*,nosep]
  \item Market research consultants
  \item Public policy advisors
\end{itemize} &
\begin{itemize}[leftmargin=*,nosep]
  \item Excel spreadsheets
  \item Social media data on X
\end{itemize} &
\begin{itemize}[leftmargin=*,nosep]
  \item Demo \& co-design sessions
  \item Customer interviews
\end{itemize} &
\begin{itemize}[leftmargin=*,nosep]
  \item Supports new ways of thinking about common business analysis tasks such as CSV labeling and sentiment analysis
\end{itemize} \\
\bottomrule
\end{tabularx}
\caption{Summary of studies, participant groups, datasets, methods, and takeaways from the \textit{Teleoscope} design process}
\Description{Table of studies, participant groups, datasets, methods, and takeaways from the \textit{Teleoscope} design process}
\label{tab:study_summary}
\end{table*}

\subsection{Formative Study}
\subsubsection{Participants}
For our formative study, the users we refer to were members of our design team and a larger set of lab members who were not involved in \textit{Teleoscope} development. In terms of expertise, all users were trained computer scientists from the upper-level undergraduate to post-doc level; some members of our team were trained UX and qualitative research practitioners from the upper-level undergraduate to PhD level. Several users we talked to had experience conducting thematic analysis in previous studies.

\subsubsection{Dataset}
As described in Section~\ref{sec:RW Qual Analysis}, we were motivated by the increasing amount of qualitative work that used Reddit as a qualitative site of inquiry into social interaction and cultural norms. Prior to working on the \textit{Teleoscope} system, members of our research team were interested in exploring social norms around account sharing within the context of Internet security and privacy. For this reason, we used Reddit's "Am I the Asshole" (r/AITA) advice forum~\cite{reddit2024aita} as the data source for our formative studies as it was both (a) one of the largest and most active sub-communities, or sub-Reddits as they are known on the platform, and (b) because it contained many interesting, explicitly normative accounts of conflicts regarding social behaviour and cultural norms (some of which had to do with our original research questions around digital security/privacy). The r/AITA corpus is also too large to be tractable for a small research team, which fit our problem domain well. It contains nearly 1 million top-level documents, or posts, which we filtered down to around 350K usable documents by removing posts that were removed by moderators or otherwise corrupted\footnote{At the moment, \textit{Teleoscope} does not support nested document display such as with Reddit's comment threads, although the interface can display URLs to the original post should the user be interested in this information.}.


\subsubsection{Methods}
We used a variety of informal low-cost UX evaluation methods to iterate on early design choices and prototypes, including cognitive walkthroughs, heuristic evaluations, and informal observations with our users. After each round of informal evaluation, we would incorporate user feedback into a subsequent design prototype. The cognitive walkthroughs and heuristic evaluations were performed with standard methods and heuristics taken from the Nielsen Norman group~\cite{nielsen90heuristics, nielsen92heuristics, Neilson}. Informal observations were performed on low-level system interactions such as menu clicking and basic keyword searches to discover and amend heuristic violations. 

In these formative stages, the \textit{Teleoscope} interface changed drastically from an interface and visualization standpoint, moving from a "top-down" approach where the entire corpus was visualized on the screen to a more "bottom-up" approach where only documents of interest were displayed on the interface. The research team would convene at least once a week to discuss results of formative prototypes and design choices. After a few months of prototyping in this manner, the research team collated a series of user metaphors and design goals for the rapidly maturing interface. We then decided to transition to a more longitudinal evaluation of the interface after the research team was satisfied the interface was robust enough and had met the established design goals to a standard that allowed for more in-depth user testing.

\subsection{Results}
\begin{figure*}[t]
    \centering
    \includegraphics[width=\textwidth]{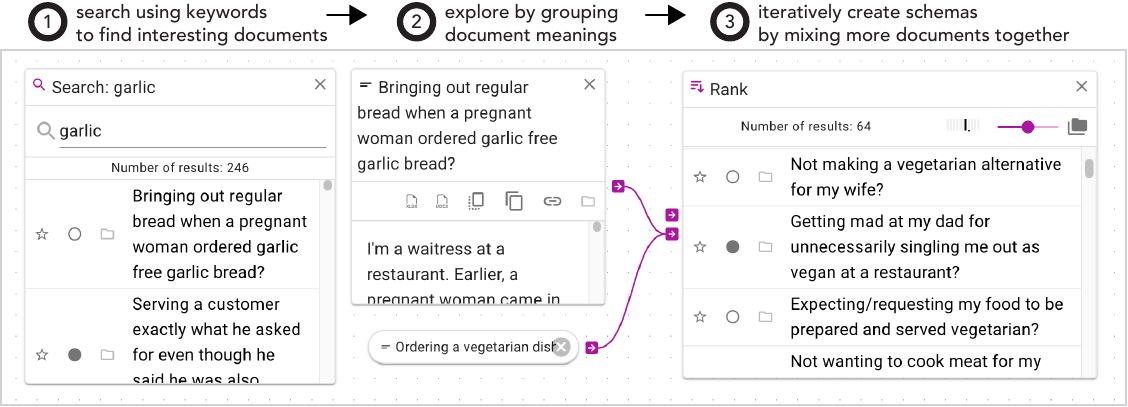}
     \caption{The core \textit{Teleoscope} workflow: Beginning with keyword searches, researchers can iteratively select and mix groups of documents together in a process we call \textit{thematic curation}, then explore NLP-based document suggestions from the larger corpus to conduct interpretive data curation and achieve information saturation on resultant \textit{thematic schemas}.}
     \Description{Diagram illustrating the three-part Teleoscope workflow of searching using keywords to find interesting documents, exploring by grouping documents with similar meanings together, and iteratively creating schemas by mixing more documents together}
    \label{fig:usage_scenario}
\end{figure*}

The results of our formative study were a set of design goals and system metaphors to support an interpretivist approach to data curation (i.e., thematic curation) on large text corpora. It also resulted in an interface for visualizing the process of conducting thematic curation and the resulting artifacts (thematic schemas). We report these results by walking through a usage scenario for \textit{Teleoscope} using examples based on the r/AITA dataset throughout to provide commentary on and trace through our design decisions throughout the initial development of \textit{Teleoscope}. The process is summarized in Figure~\ref{fig:usage_scenario} and additional usage scenarios provided in Figure~\ref{fig:teleoscope-process}.

\subsubsection{Keyword searches: Schema nucleation}
Our original prototypes for \textit{Teleoscope} used dashboard metaphors \cite{lee2021understanding,terragni2021,gad2015themedelta,choo2013utopian,kim2020architext} where each module displayed elements such as included or excluded keywords, document similarity statistics, and topic clusters. Initial corpus visualization attempts drew inspiration from common visualization solutions such as weighted adjacency matrices of keywords and documents \cite{low2020natural, behrisch2016matrix, arruda2022vosviewer}, but due to the size of the corpus, screen pixel overlap quickly made this approach intractable. In addition, our users felt the relationship between the different dashboard modules were disjointed from the way they conceptualized data. Instead, many users who had experience in conducting qualitative work on large text corpora told us they tended to begin their workflow by quickly skimming through the dataset in order to begin building up mental "schemas".

Probing further, we learned that these mental schemas were similar to the concept of schemas in psychology, which refer to a set of beliefs, attitudes, and emotional responses related to an archetypal situation \cite{berret2022iceberg}. Berret and Munzner appropriated the concept of schemas in their work on sense-making processes with data visualization \cite{berret2022iceberg} by arguing that data visualization provides a method by which a user can externalize internal schemas, allowing them to inspect, amend, and refine internal schemas alongside externalized artifacts of the schema. For our users, this process of schema "nucleation" had to begin with some entry point into the dataset. The most common approach was either by skimming documents or keyword search, as many of our users with qualitative research backgrounds had formal training in methodologies for managing keyword searches, such as boolean search strategies \cite{aliyu2017efficiency}. 

They also explained that every keyword search or pass at the data allowed them to iteratively subset, facet, or create "views" on the data. This is similar to the process of \textit{crystallization} described by Ellingson \cite{ellingson2009engaging}, where triangulation of a phenomena many times results in more nuanced interpretations of the data, akin to a more complex crystalline structure. This insight also supports a nice metaphorical similarity to graph visualizations that informed our subsequent design of \textit{Teleoscope's} data structures, which we expand on in Section~\ref{sec:Implementation}. As a result, \textit{Teleoscope} supports standard fuzzy keyword search operations and windows for dataset skimming to enable schema nucleation. We felt these were suitable user metaphors for initial corpus exploration due to user familiarity. After these conversations, we also abandoned the dashboard metaphors in favour of a windowing system with visible input/output areas on a non-linear canvas interface. By moving commands and system state out of menus and collapsible dashboard modules, we felt we were able to better provide a feeling of working "close to the data". A design with a minimum display of information was also able to better facilitate drawing connections between groups of documents that were user-defined.

\subsubsection{Document Grouping: Initial Forays into Curation}
Using our new prototype, users explored the r/AITA dataset for privacy/security conflicts using keyword searches. This quickly raised another important question from our users: How did the user know they had the right keywords? Concerns about biasing the curatorial process through the choice of inappropriate or \textit{a priori} keywords arose. Another drawback of using keywords was the lack of contextual differentiation. For example, using the keyword \textit{privacy} brought up relevant documents, such as those about changing passwords following an account sharing conflict, but also less relevant documents in the corpus, such as using the bathroom in private. We realized the interface needed to provide a mechanism to annotate, group together, and differentiate relevant and irrelevant documents to the researcher. In this initial phase of curation, these document groups often captured vague ideas about under-determined concepts in the dataset and the researcher's mind. Although this interactive process helped researcher's refine ideas about their internal schemas, they did not yet have strong commitments to the external schemas represented in the interface.

Eventually, as the concepts in document groupings became clearer to the user, the interface became a visual artifact, or externalization of the ambiguous internal schema. At this point, the interface could often become cluttered, with many annotated documents "in staging" with no clear meaningful arrangement. We therefore implemented an action history system where every system action was logged and could be rolled back. We opted for an algebraic workflow metaphor to allow the user to directly manipulate the history of their actions on the interface via the windowing elements.

At this stage, users might be curious about whether their initial document groupings represented meaningful, substantiated patterns in the larger text corpus and not just isolated examples. To allow users to discover this information, we experimented with various NLP approaches that allowed users to make sense of their keyword searches and initial groupings at scale. We settled on using vector similarity search provided by LLM embeddings as an approximation of semantic similarity (see also Section~\ref{sec:RW AI}) within a group of documents. Although vector similarity is not a perfect analogue to the thematic similarity being developed in the user's mind, since the former is statistical and the latter interpretive, our users reported a sense of interactive, serendipitous discovery that the research team found satisfactory and wanted to substantiate throughout the data curation process.\enlargethispage*{16pt}

\subsubsection{Information Saturation in Corpus Construction}
By leveraging the NLP features in \textit{Teleoscope}, users were now able to iteratively refine their document groupings, receiving automated suggestions on-demand for new documents to add to their document groups. These changes would subsequently propagate to our backend in order to provide increasingly user-driven suggestions based on the vector relatedness of documents users placed in the same group (see Section~\ref{sec:Implementation} for system details). Eventually, we hoped that users would feel a sense of information saturation in their curation process. Richness and saturation are two concepts used to discuss rigour during qualitative analysis \cite{mcdonald2019reliability}. In contrast with quantitative research paradigms, rather than valuing many data points (i.e., through statistical power), qualitative researchers are particularly interested in having many \textit{unique} data points (i.e., through information power \cite{tobin2004methodological}). As users continued to curate their thematic schemas, we observed there came a point when new documents surfaced by the \textit{Teleoscope} system did not provide new insights into a particular phenomena the researcher was interested in, indicating thematic curation was complete.

Determining when this point has arrived is one of the chief difficulties of working with large text corpora, since it is often unfeasible, and sometimes impossible, to read every relevant document within a dataset. As mentioned in Section~\ref{sec:Related Work}, ad-hoc sub-sampling or data partitioning strategies are often employed as a result. With \textit{Teleoscope}, the NLP features were able to group together large numbers of thematically similar posts for the researcher to examine. We realized during our formative study that there were three archetypal kinds of document groups: (1) Groups of documents that could be analyzed in "batches" since they contained thematically similar content; (2) groups of documents that could be confidently ignored by the researchers since they were not thematically relevant; and (3) special example posts that we called \textit{signposts}. Signposts were nuanced "exceptions that proved the rule" that warranted further interpretive focus and attention.

For example, one user read a post titled "AITA for secretly muting my wife's emails while on vacation?". The researcher's interpretation of the post was that there existed conflicting value systems at play. On the one hand, the husband was clearly violating a privacy norm by altering his wife's email settings without her permission. On the other, he seemed to believe that he was doing her a favour, and quite possibly improving their vacation experiences. This signpost, along with other signposts, helped the researcher to solidify thematic curation around the concept of "ambiguous privacy/security violations that challenge the accepted value systems of current privacy models". Needless to say, this is an extremely difficult concept to search for using keywords, and requires carefully reading through most of a NLP-generated document group provided by the system. This is because the content of the post is a nuanced problem about digital privacy, which is rarely surfaced by a statistical vector similarity search. This example helped inform the fundamental difference between what a system could easily do with a vector search versus the value of human interpretation of the data.

Our saturation goal therefore became to find the largest number of thematically unique examples of nuanced situations such as these. Semantic vector similarity search gave us a recursive process, allowing researchers to partition datasets into large groups of documents that represented the same thematic concept, documents that were not thematically relevant, and special signposting examples. Based on our formative study, we determined that saturation had been achieved when further faceting could not meaningfully differentiate groups of documents; as a corollary, saturation had occurred when we could not find any further examples that could signpost differentiating aspects in our thematic schemas.

\subsection{Design Goals}
Having surfaced the usage metaphors that needed to be supported in a qualitative approach to data curation through our formative study, the research team collated them into a set of design goals based on our observations of the data curation process with the r/AITA dataset. 

Our design goals target live, qualitative interactions (\textbf{DG1}) with data, as opposed to statistical topic modeling packages that run as terminal-based Python programs~\cite{scikit-learn, rehurek2011gensim, devlin2019bert, el2019semantic}. We also visualize the \textit{process} of thematic data exploration and curation so that collaborators can share their thought processes and interpretive conclusions (\textbf{DG2}). This is different than topic modeling visualization systems that focus on high-level statistics, topic hierarchies, word counts, and labels~\cite{el2019semantic, kim2019topicsifter, terragni2021}, which we discarded after completing our formative study. To support our downstream target users --- who were trained in qualitative methods but not computer experts --- we created a web platform to facilitate "fast enough" interaction so that target users could feel as if they were having a creative and improvisational experience with the data (\textbf{DG3}). 

To summarize, our design goals for the system were:
\begin{itemize}[leftmargin=*]
    \item \textbf{DG1. Imbue large-scale data curation with familiar qualitative interactions} Qualitative researchers are familiar with \textit{arranging}, \textit{annotating}, and \textit{grouping} of documents physically in their data sense-making. We borrow from these interaction paradigms and imbue them with computational power to enable interpretive data curation with large-scale datasets.
    
    \item \textbf{DG2. Help researchers externalize their internal schemas as visual artifacts that can be \textit{retraced}, \textit{modified}, and \textit{inspected} by collaborators.}
    When curating data, capturing the steps of the interpretive thought process through our action history is important in fostering collaboration with team members, who can inspect, share and modify workflows and understandings of the corpus.
    
    \item \textbf{DG3. Facilitate an on-going feeling of interpretive, creative, and serendipitous sensemaking for researchers.} The feeling of continuous interactivity was important for us to facilitate through our low-level system design choices in order to foster a sense of interpretive immersion and to avoid letting users get creatively blocked waiting for long computations on the entire corpus.
\end{itemize}

\section{System Specification}
\label{sec:System}
\begin{figure*}[t!]
    \centering
    \includegraphics[width=\textwidth]{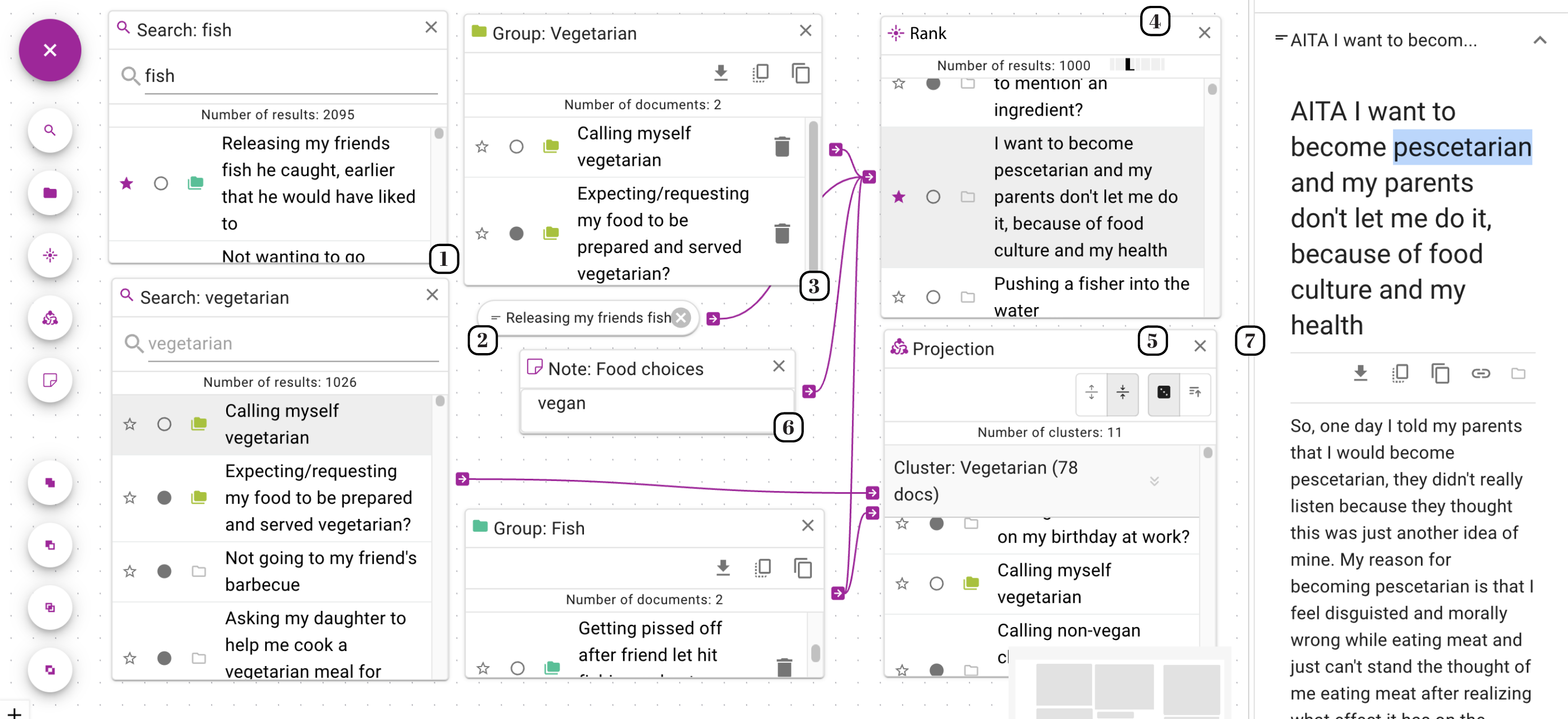}
    \caption{An image of the \textit{Teleoscope} workspace. (1) Users start by performing a \textbf{keyword search} to explore documents using the Search node; (2) \textbf{Documents} are dragged onto the workspace; (3) Documents can be put into \textbf{Group} nodes for organization; (4) \textbf{Rank} nodes can use documents, Notes or Groups as control inputs; (5) \textbf{Projection} nodes create clusters using Groups as control input; (6) \textbf{Note} nodes contain arbitrary text which is also vectorized and can be used as an input to Rank nodes; (7) the \textbf{sidebar} has a quick viewer for documents, saved items, bookmarks, and settings. Keyboard navigation is used for quick exploration and group creation.}
    \Description{Image of the Teleoscope workspace, showcasing all of the features capable on the interface: Including keyword searches, node dragging, Note and Group control inputs, Projection nodes, and a sidebar.}
    \label{fig:workspace}
\end{figure*}

In this section, we outline the system features and implementation of the \textit{Teleoscope} system. An overview of the system features is presented in Figure~\ref{fig:workspace}. The word ``Teleoscope'' is a portmanteau of the Greek word \textit{teleology} and telescope. \textit{Teleoscope} is a way for researchers to interpretively explore and curate a large, intractable corpus in an opinionated fashion, resulting in thematic schemas, groups of documents from the overall corpus that have been sufficiently "downsized" to be amenable downstream to canonical approaches in qualitative analysis, such as reflexive thematic analysis \cite{braun2012thematic} or grounded theory \cite{muller2014curiosity}. Thematic schemas can be thought of as occupying the space on a spectrum between topic categories on one end (which usually have little to no interpretive input from a human in their creation process) and the final data examples or quotations (which are used in a report or publication to substantiate a theme) in a qualitative/empirical work on the other. 

\subsection{System Features}
\label{sec:Features}
In \textit{Teleoscope}, the act of data curation takes place within a node-based visual data-flow editor. Users are initially presented with a blank, infinite canvas workspace, similar to those seen in whiteboarding applications such as Miro \cite{miro}. Users are then able to create individual nodes that represent key operations in the system, such as keyword searches or NLP-powered recommendation. Nodes can be arbitrarily arranged on the interface via drag-and-drop operations. The basic interaction premise is that data \textbf{source} nodes flow into \textbf{targets} (other nodes, see also Figure~\ref{fig:teaser}). Data sources are arbitrarily-long document lists that subset the full corpus. The \includegraphics[height=1em]{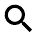} \textbf{Search} node (Figure~\ref{fig:workspace}-1) allows the full dataset to be filtered based on a fuzzy keyword search (along with basic boolean text search operations). \includegraphics[height=1em]{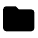} \textbf{Group} nodes (Figure~\ref{fig:workspace}-3) are document lists that contain documents users have explicitly selected for inclusion.

The search and group nodes can also be used as input to control NLP operations. If a user creates a group of documents, it can be fed into the \includegraphics[height=1em]{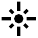} \textbf{Rank} (Figure~\ref{fig:workspace}-4) node by connecting the Group node to a target Rank node. The Rank node will then perform a vector similarity search of the full corpus based on the average of the vectors in the Group node and return an ordered list of the most similar documents up to a user-defined threshold. By default, this is an equal weighting of the documents in the Group node, although both documents and groups housed within groups can have their weights increased to emphasize certain documents during a re-rank operation. Negative examples were not needed as users could directly remove collections of the corpus that were not of interest using the regular Group mechanisms. Feedback dynamics are not directly exposed to the user. If a specific data source, like the results of a Search node, are connected, then the Rank node will only order that source. Custom user annotations, represented by \includegraphics[height=1em]{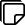} \textbf{Note} nodes (Figure~\ref{fig:workspace}-6), can also be used as input to Rank nodes. Note nodes are arbitrary text that the user enters which are immediately vectorized. A creative way to use Note nodes is to "take on the voice of the documents" to create "synthetic" archetypal data examples. These synthetic examples can then be used as an input to the Rank node in order to explore whether these synthetic data examples are actually substantiated by real documents in the corpus. This is a novel search mechanism based in creative arts practices that has essentially not been enabled with prior technology.

In addition, \textit{Teleoscope} supports automatic document clustering operations through the \includegraphics[height=1em]{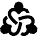} \textbf{Projection} node (Figure~\ref{fig:workspace}-5). Projection nodes take data sources and create low-dimension projections of the source documents by instantiating a custom distance metric. Technically, the "metric" is a function over a matrix customized by the user by using documents within the data source as the input to the function, we use the term "metric" to be consistent with UMAP terminology \cite{umap_embedding_space_custom_metric}. For each Projection node, the custom distance metric is created by calculating the minimum distance between each document within the data source and a maximum distance between documents in other data source nodes currently on the interface. This custom metric is then used to cluster the remaining documents in the full corpus. In practice, this means that the Projection node attempts to group all the documents in the corpus as if they were in the same Group nodes that the user has defined. In this way, \textit{Teleoscope} supports the archetypal interactions of a qualitative researcher when curating documents such as \textit{arranging}, \textit{annotating}, and \textit{grouping} while enabling this work at scale via NLP operations that allow the corpus to be quickly explored and faceted.

\textit{Teleoscope} also supports collaborative inspection of workspaces, allowing members of a research team to share their workspaces with one another in order to inspect, trace, and re-compute \textit{thematic schemas}. Team members can synchronously share workspaces with one another, allowing them to make all the same actions as they would on their own workspaces, including moving nodes, commenting, and comparing schemas. Workspaces can also be cloned. This allows members of a collaborative research team to experiment with each other's thematic curation processes as digital artifacts and surface discussions about curatorial decisions made over the course of refining the corpus.\enlargethispage*{16pt}

\subsection{Implementation}
\label{sec:Implementation}
\subsubsection{Visualization Interface}
To implement our visualization interface, we used ReactFlow, a React library for constructing node-based editors on top of NextJS. Interface state is managed through middleware that routes to a MongoDB database, allowing us to inspect user action histories, although these were not directly exposed to users in our evaluations.



\subsubsection{NLP Operations}
\textit{Teleoscope} can be "hot-swapped" for any off-the-shelf exogeneous model, such as those available on the HuggingFace repository \cite{wolf2019huggingface}: In our user evaluations, \textit{Teleoscope} uses FlagEmbedding's BGE-M3 model \cite{bge_m3}. We use Group nodes (a collection of user-specified documents) as control inputs to define a similarity metric (i.e., the "supervision" portion of our semi-supervised approach). The similarity metric situates documents within the same Group node near 0 and are normalized to 1 between Group nodes. As mentioned in Section~\ref{sec:Formative Study}, users are also able to create their own examples of synthetic data and add these to document Groups to verify hypotheses about the kinds of data contained in the corpus. We use Universal Manifold Approximation and Projection (UMAP) for reduction to five dimensions~\cite{mcinnes2018umap-software} with hyper-parameters \verb|n_neighbors=100| and \verb|min_dist=0.05| . The Projection nodes that provide clustering functionality are implemented using Hierarchical Density-Based clustering (HDBSCAN*)~\cite{mcinnes2017hdbscan} with hyper-parameters \verb|min_cluster_size=5| and \verb|min_samples=5|.

\section{Evaluation \& Deployments}
\label{sec:Evaluation}
This section outlines our evaluation processes and results. In addition to the formative study described in Section~\ref{sec:Formative Study}, we evaluated \textit{Teleoscope} by conducting (1) a second, multi-week formative study using a "simulated research team" of trained qualitative researchers (N=5), followed by a post-hoc review of our interface and data with a group of visualization experts; (2) a multi-month field deployment within a qualitative research group; and (3) an on-going public release as we continue developing the tool. Evaluations are summarized in Table~\ref{tab:study_summary}. We opted to conduct a second formative study as part of our evaluation as we found it difficult to make informative baseline comparisons between \textit{Teleoscope} and other interactive data exploration tools. This was due to the order of magnitudes difference in the scale of data typically being ingested and manipulated by \textit{Teleoscope} versus other existing tools used by qualitative researchers, and the manner in which the data was handled at the document rather than corpus level (we expand on this in Section \ref{sec:Discussion}). Although the number of participants in these deployments was relatively small, we followed their use of the system for a much longer period of time than in a traditional system evaluation. Part of the motivation for this was our aim to understand and support how qualitative researchers actually got work done, and also in light of the fact that we were not comparing \textit{Teleoscope} to some experimental baseline system. It was therefore paramount for us to observe whether \textit{Teleoscope} was relevant to our target user population (in line with RtD evaluation criteria) to evaluate whether the system was successful. User studies were conducted under the approval of our institution's research ethics board; we obtained participant consent forms and reminded participants of their right to cease participation in the studies if desired.

\subsection{Extended Formative Study with "Simulated" Research Team}
Once our system implementation had stabilized, we released the first version of \textit{Teleoscope} for a second, multi-week formative study with representative target users. The premise was to simulate a research team working on the same research question with a large text corpus that we provided. We were interested in the following research questions:

\begin{itemize}[leftmargin=*]
    \item \textbf{RQ1}. In what ways did participants understand and use \textit{Teleoscope's} system features such as Search, Rank, and Projection?
    
    \item \textbf{RQ2}. How did participants incorporate \textit{Teleoscope} into their understanding of qualitative research processes?
\end{itemize}

We were also interested in observing the extent to which \textit{Teleoscope} could hold up in a simulated production environment and welcomed ongoing bug reports from participants. We therefore had to ensure the interface was robust enough for participants to use on their own devices, outside of a lab environment.

\subsubsection{Participants}
We recruited participants who were familiar with qualitative methods: at least an upper-level undergraduate course and/or equivalent research experience. Eight participants were recruited; three dropped out before the multi-week study was set to conclude (final count $n=5$). Of the participants who remained, one was a PI at a university who led qualitative research in the School of Nursing; one was a senior PhD student in Sociology; two were upper-level undergraduates in Sociology working on self-directed projects in a qualitative methods course; and one was an upper-level undergraduate in Psychology who had experience with qualitative work. Participants were reimbursed at a rate of twice the local minimum wage owing to their status as expert users.

\subsubsection{Dataset}
The topic that was chosen by study participants for investigation was \textit{Critical and end of life care} within the \textit{r/AmItheAsshole} dataset (see Section~\ref{sec:Formative Study}). The group brainstormed seed keyword searches such as \textit{Medical Assistance in Dying (MAID)}, \textit{End of life care}, \textit{Palliative}, \textit{ICU}, \textit{Failures}, \textit{Emergency rooms}, \textit{Emergency care}, \textit{Lack of beds}, and \textit{Overcrowding}. 

\subsubsection{Methods}
Participants were introduced to \textit{Teleoscope} and each other during a one-hour training session where we brainstormed a shared research topic. Participants were then instructed to use \textit{Teleoscope} for at least 10 hours before a focus group scheduled three weeks later. For each session that they used the system, they wrote a diary entry detailing (1) the \textit{concepts} that they explored; (2) the \textit{process} by which they explored the theme; (3) any \textit{collaboration} features they used; and (4) any bugs or features requests. No significant interface changes were made during the course of this extended formative study, but minor bugs were fixed. System logs were kept throughout the study, which we verified to confirm participants used the system for a minimum of 10 hours before the focus group.


\begin{figure*}[t]
    \centering
    \includegraphics[width=\textwidth]{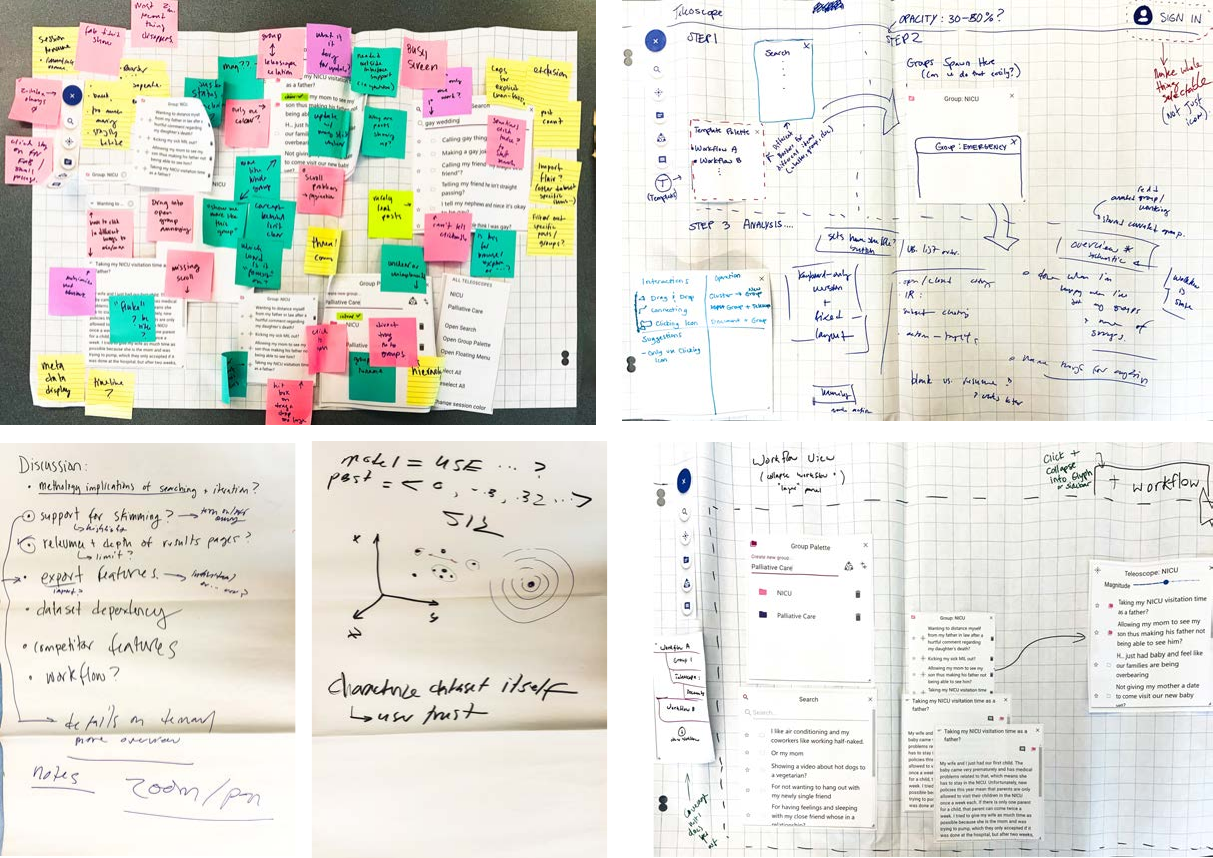}
    \caption{Large printouts of the \textit{Teleoscope} interface were used to draw and annotate problems and design ideas with both the focus group (left), and the visualization group during a post-hoc analysis (right).}
    \Description{Multiple large printouts at various stages of the Teleoscope development process that were annotated by participants.}
    \label{fig:focus-group-infovis-results}
\end{figure*}

We collected the diary entries and analyzed them using affinity diagrams before the participant focus group, which took three hours, including a lunch break. The focus group involved (1) diary discussion; (2) brainstorming on problems encountered; (3) brainstorm on feature requests and design solutions to identified problems; and (4) explanation of ML concepts unfamiliar to participants and brainstorming on better alignment between visual and interaction metaphors. We video and audio recorded the focus group and used large printouts of the \textit{Teleoscope} interface to draw and annotate problems and design ideas (see Figure ~\ref{fig:focus-group-infovis-results}).

\textit{Post-hoc Analysis by Visualization Group}
After we had analyzed and summarized participant results, we presented \textit{Teleoscope} in two multi-hour analysis sessions with a visualization research group. Our results reflect the analysis of that group along with our own analysis and solution brainstorming. 

\subsubsection{Results}
\textit{Diary Results}. \textbf{Using \textit{Teleoscope} differed from keyword search paradigms.} Participants found that the Rank operation differed significantly from keyword search operations they had previously used. For many participants, it took some time to (1) differentiate the output of a keyword search from the output of a Rank operation; and (2) to differentiate valid results that did not meet their expectations due to the documents that existed in the corpus from invalid results due to bugs in the system or mental model inconsistencies. There were negative transfer effects from prior familiarity with keyword searches that we noticed took participants multiple sessions to unlearn. 

For example, P1 searched for ``palliative'' and was \italicquoteinline{...surprised by how many posts were about animals at end of life, which does not fit our defined research topic.} P1 then wished \italicquoteinline{...there was a feature that would take everything I had already put within one group and give me `more like this'}, which was exactly what the Rank operation was designed to do. Multiple participants corroborated this sentiment in their diaries (P5, P7, P8). This indicated problems in the participants' mental model, likely due to (1) how we were representing the Rank operation on the workspace; (2) our training, documentation and support materials; and (3) not enough time to learn the tool.

However, P1 reported for their third session that they spent a long time looking through the documentation and support videos the research team developed to understand the possibilities of \textit{Teleoscope}: 

\aptLtoX[graphic=no,type=html]{\begin{quote}
Today I also spent time trying to go down the rabbit hole of different searches to try to gain a true appreciation for how this machine learning approach to data collection differs from just keyword searching within the Reddit search. This was really evident to me when I found a post where the OP had a palliative/terminal illness, and I wanted to find others where this was the case. I made a new folder for this category, then used the [Rank] feature, and immediately found one other post where the OP has cancer and was asking a friend to not mourn their death. It would be extremely difficult to keyword search for this type of topic, but it’s a very interesting and important area to capture (OPs with terminal illnesses). This was a great exploration! (P1)
\end{quote}}{\begin{italicquote}
Today I also spent time trying to go down the rabbit hole of different searches to try to gain a true appreciation for how this machine learning approach to data collection differs from just keyword searching within the Reddit search. This was really evident to me when I found a post where the OP had a palliative/terminal illness, and I wanted to find others where this was the case. I made a new folder for this category, then used the [Rank] feature, and immediately found one other post where the OP has cancer and was asking a friend to not mourn their death. It would be extremely difficult to keyword search for this type of topic, but it’s a very interesting and important area to capture (OPs with terminal illnesses). This was a great exploration! (P1)
\end{italicquote}}

This indicated that it was possible to learn the difference between keyword search and Rank operations, but that the learning curve was steep enough to require multiple hours of usage and documentation review. 

\textbf{\textit{Teleoscope} can support quick, iterative curation strategies.} P7 articulated a very clear document review strategy and seemed to understand the tool very quickly. Ignoring the group topic, they began searching for documents related to their own research program by skimming titles:

\aptLtoX[graphic=no,type=html]{\begin{quote}
I was interested in [AITA] posts about gay marriage, which is a topic tangentially related to my own research. I populated the [gay wedding] group with results whose titles caught my eye. I should note that I very rarely read the actual documents. If the title was vague, I occasionally skimmed the first few lines. (P7)
\end{quote}}{\begin{italicquote}
I was interested in [AITA] posts about gay marriage, which is a topic tangentially related to my own research. I populated the [gay wedding] group with results whose titles caught my eye. I should note that I very rarely read the actual documents. If the title was vague, I occasionally skimmed the first few lines. (P7)
\end{italicquote}}

They further suggested adding a document quick viewer to aid in skimming. They then organized documents into groups, re-labelling and changing the groups as they developed their understanding of the corpus and the tool. Then, they switched between using the Rank operation and the group feature to find relevant documents:

\aptLtoX[graphic=no,type=html]{\begin{quote}
After adding about half a dozen documents to the group `gay stuff', I noticed that many of the documents are about gay panic. That is, the fear of being (wrongly or correctly) perceived as gay, the dislike of anything perceived as gay, and an aversion from being around gay people. I changed the group name to `gay panic` to reflect this...once I had about 13–14 results, I opened the [Rank] window for the group. Looking at the first two pages of results, none of them that weren’t already in the group seemed very relevant, mostly judging by their titles and occasionally by the first couple of sentences in the document. I refined the search...[with a] couple of documents that I thought particularly demonstrated gay panic. (P7)
\end{quote}}{\begin{italicquote}
After adding about half a dozen documents to the group `gay stuff', I noticed that many of the documents are about gay panic. That is, the fear of being (wrongly or correctly) as gay, the dislike of anything perceived as gay, and an aversion from being around gay people. I changed the group name to `gay panic` to reflect this...once I had about 13–14 results, I opened the [Rank] window for the group. Looking at the first two pages of results, none of them that weren’t already in the group seemed very relevant, mostly judging by their titles and occasionally by the first couple of sentences in the document. I refined the search...[with a] couple of documents that I thought particularly demonstrated gay panic. (P7)
\end{italicquote}}

\textbf{Positive and negative transfer effects from other qualitative research software.} Participants' extensive prior experience with qualitative research software allowed them to have a much clearer mental model of the tool without extensive training, which indicates the possibility of positive transfer effects. Many desired features were given as examples from tools that they had experience with, such as Google Docs, MaxQDA, and NVivo. Unsurprisingly, they mostly expected \textit{Teleoscope} to work like other interfaces they had previously used. To our surprise, none of the participants used any of the collaboration features embedded in the system during this deployment.

P7 noticed many problems with the interface and made suggestions for design changes that were brought to the focus group, including (at this stage), a lack of annotation and data coding features, document export features, and overall corpus visualization features: \italicquoteinline{At this point of my exploration of this theme, being able to play around with how [documents] connect to one another [...] 
might have helped refine my thinking.} Participant diary results were summarized and presented at the focus group, motivating our central discussion points during the session.

\textit{System Log Results}. During the user study, the system maintained a log of user actions as they interacted with the system. Across the 5 participants, 7 sessions with the Rank operations were tracked (one participant created 3 separate sessions for themselves). The mean number of actions tracked per session was approximately 310 (median = 286, minimum = 97, maximum = 684). Actions included such things as session initialization, creation/movement/deletion of nodes, keyword searches, instantiation of ranking and results, in short, any conditions where the state of the workspace was altered.

The actions were logged and then visualized to better understand user interactions. Overall, users made use of an iterative process to find documents of interest, alternating between putting documents into groups and instantiating new Rank operations to find new documents relevant to their queries and then sorting them into their groups. The number of Rank operations created across the study were however, relatively small (between 2 and 4), but indicated that the iterative curation strategy supported by \textit{Teleoscope} showed promise for interpretivist-style curation.

\textit{Focus Group Results}. The focus group provided insight into the needs of qualitative researchers with different levels of expertise with computer supported analysis. We used a thematic analysis approach to review the focus group results. We report here on the most prominent themes from our analysis.

\textit{Mental model confusion regarding the limits of system operations}. 
In the system deployed to the participants of our extended formative study, Rank nodes were originally coupled directly to a singular Group node. When documents were added to the Group node, the changes were propagated to the Rank node. This led to confusion about the state of the Rank node and participants expressed a desire for more explicit visual representation of the Rank operation and the ability for more fine-grained control of the Rank operation, for example by allowing multiple input sources into the Rank node. Participants also wanted some visualization support for understanding how they were \italicquoteinline{going down different paths} (P7) in their curatorial process. Participants agreed that \italicquoteinline{seeing it} and \italicquoteinline{understanding how close documents are to each other in space [is important].} (P1). In addition, since participants were most familiar with and had prior training in keyword manipulation, set operations were desired to give users confidence that they were being thorough in their curation process.

\textit{Concerns over the methodology supported by \textit{Teleoscope}.}
Participants agreed that they did not need to show total path \textit{exhaustion}, but instead a sense of \textit{saturation} (see Section~\ref{sec:Formative Study}) and demonstrable information power. One issue with this is that although participants had no issues with the \textit{Teleoscope} process, they expressed concerns about how to explain the \textit{Teleoscope} process to reviewers at a high level. Some participants shared experiences with defensive publication strategies: in some qualitative papers, keyword searches are reported on directly to demonstrate that they had reached saturation or sufficient information power within their dataset. \textit{Teleoscope's} radically different approach to document curation introduced concerns around methodological reporting, especially since there was no clear literature to point reviewers to that demonstrated the efficacy of \textit{Teleoscope's} curation strategy (see Section~\ref{sec:Discussion}).

Participants were also unsure whether using \textit{Teleoscope} constituted a violation of rigour in that it seemed to mix the data curation and analysis phases in a qualitative research lifecycle. If text-level analysis in particular was supported, some participants felt that it might not be appropriate to allow for both aspects to be supported in the interface at once. Despite this, we found that participants used \textit{Teleoscope} as we had intended: they explored data in a manner that extended beyond keywords searches. For example, participants reported the following: 

\aptLtoX[graphic=no,type=html]{\begin{quote}
I was excited when the tool came up with things about the topic, but not including keywords that I used. (P1)
\end{quote}}{\begin{italicquote}
I was excited when the tool came up with things about the topic, but not including keywords that I used. (P1)
\end{italicquote}}

\aptLtoX[graphic=no,type=html]{\begin{quote}
I use it to find [documents] that I wouldn't have found. (P7)
\end{quote}}{\begin{italicquote}
I use it to find [documents] that I wouldn't have found. (P7)
\end{italicquote}}
This led us to believe that \textit{Teleoscope} could be used for a longer and more in-depth research project, which motivated our subsequent field deployment.

\subsubsection{Post-Hoc Analysis and Recommendations from Visualization Group}
We presented our results and current interface to a visualization group at our institution. After two multi-hour analysis sessions, the visualization group recommended the following:

\textbf{Visualize workspace relations}. Up until this point, our interface did not include any visual graph concepts since our original abandoning of adjacency matrices. The recommendation was to clarify system state further by treating windows as nodes which could be used as input sources for operations.

\textbf{Clarify data type representation}. Since the results of workspace operations were effectively variations on ordered lists, the recommendation was to create visual homogeneity where data types were the same, and visual differentiation where they were different. From this, we developed our current paradigm of source, control, and target data types, and a unified output datatype of a list of documents.
    
\textbf{Create a quick-viewer to allow for minimized windows}. A sidebar was recommended to allow documents to remain as minimized pills on the interface while users reminded themselves of document contents.
    
\textbf{Allow for multiple workflows}. Users wanted to simultaneously pursue multiple explorations with quickly accessible workflows. They further recommended that we create workflows that could be chained to target both provenance and re-usability.
    
Our main takeaways from the visualization group recommendations were to move our original window metaphor into a node-based metaphor with simplified workspace objects. This required a major redesign of both our front and backend systems to support graph operations. It also introduced questions of concurrency and graph directionality. For example, our first iterations of this design introduced cycles into the graph; as such, we made the decision to make sources strictly ``left-handed'' and targets ``right handed'' with the exception of set operations. This had further impacts on our state management system, which had to be redesigned to work with a graph-based structure.

\subsubsection{Extended Formative Study: Conclusion}
In this deployment, we investigated how real qualitative researchers would use \textit{Teleoscope} in a simulated research environment. Coming back to our research questions and summarizing our findings:

\begin{itemize}[leftmargin=*]
    \item \textbf{RQ1 Result}. Participants understood the overall concept of the Rank operation and the workspace as a whole, likening it to walking down different paths in an unknown region. Even though this metaphor came from the participants, the operation inputs and system state were often unclear. We decided that the path metaphor should therefore be visualized as directly as possible. Participants did not use any collaboration features of \textit{Teleoscope}, which may be due to a lack of impetus, our study environment, or because of our design. In the next evaluation, we addressed these possibilities to ensure collaboration features were used.

    \item \textbf{RQ2 Result}. Participants interpreted \textit{Teleoscope} primarily as a data curation tool. Further, they were unsure about whether it infringed on the analytic phase of qualitative work and had concerns about academic rigour if it did.
\end{itemize}

\begin{figure*}
    \centering
    \includegraphics[width=\textwidth]{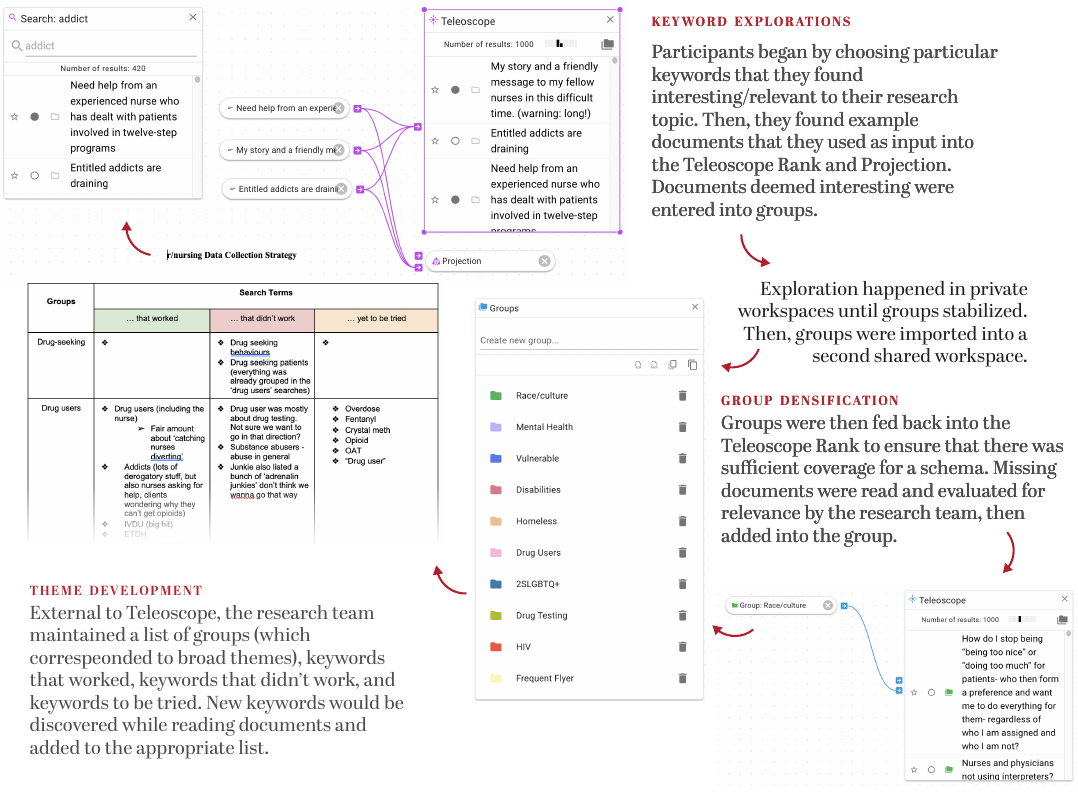}
    \caption{An example of a workflow from our long-term field deployment. Pictures are actual screenshots of research artifacts from our participant research team's data curation phase. Participants worked both individually and collaboratively on the \textit{Teleoscope} interface, and collaboratively on Google Docs, Zoom, and in-person. Due to the existing qualitative research culture in their research group, keyword searches were the focus of their data curation approach. ML features were used to discover new keywords, find thematically similar documents that did not have specific keywords, and to saturate groups with relevant documents.}
    \label{fig:allie-workflow}
\end{figure*}
\subsection{Field Deployment}
After completing an additional system redesign after our second formative study with the simulated research team, we were interested in a focussed, long-term field deployment with real qualitative research groups who could bring their own research needs to us. In a similar vein, we wanted to move \textit{Teleoscope} out of a simulated environment and into a production environment where real-world motivations and difficulties would be encountered. Our extended formative study had many artificial contrivances, such as a research topic that none of the participants were using for their own research. We felt it was important to see how \textit{Teleoscope} would perform when participants were subjected to all the benefits, consequences and costs of real research. We also committed to providing ongoing feature development and bug fixes for the field deployment, drawing inspiration from co-design methodologies in HCI. 

For the field deployment, we were interested in the following research questions:
\begin{itemize}[leftmargin=*]
    \item \textbf{RQ3}. When put in a real-world environment, how do researchers incorporate \textit{Teleoscope} into their research practices? 
    \item \textbf{RQ4}. Using \textit{Teleoscope}, to what extent are researchers able to feel confident that they were able to retrieve data using criteria that are important to qualitative researchers, i.e., richness, information power, saturation, etc.?
\end{itemize}

During this period, we also moved \textit{Teleoscope} out of a test environment into a publicly-available production release. This involved adding many security features and backup systems which were exposed to the threat of arbitrary internet attacks. We had at least one successful security breach which was dealt with and mitigated; the number of unsuccessful attacks were unknown.

\subsubsection{Participants}
Three PIs and research groups were recruited; only one research group was able to commit to long-term use of \textit{Teleoscope}. The final team comprised the PI from our extended formative study and two of their graduate RAs that were recruited specifically to use \textit{Teleoscope}. Participants were not reimbursed for their time by us since they were being supported extensively by our design team for their own research projects; we understand that the RAs were compensated in accordance with their regular duties by their PI.

\subsubsection{Dataset}
The research team used \textit{Teleoscope} to perform data curation for a research project on nurses and structural inequality as articulated on Reddit's \textit{r/nursing} sub-Reddit, a forum for working nurses to post about their day-to-day problems. On their own initiative and in alignment with their own qualitative methodological practices, the team's main conceptualization of data curation was via an external Google document that contained a set of keywords to be used in searches (see Figure~\ref{fig:allie-workflow}). After exploring the data using a variety of features in the \textit{Teleoscope} workspace (described below), they would update their keyword list with successful and unsuccessful searches. 

\subsubsection{Methods}
\textit{Teleoscope} was deployed in a standard beta release manner where participants were given a private link to \textit{Teleoscope} until we transitioned to a full public release. Participants were invited to participate in a Discord server to make bug reports and feature requests. Depending on the request, occasional emails and video interviews were conducted. Logs were kept of system use to compare with study results. For the purposes of this paper, we "terminated" data collection after six months, but their usage continues to be supported by the research team.

\subsubsection{Results}
Our field deployment helped us confirm and discover the following insights:
\noindent
\textbf{\textit{Teleoscope} helped discover unknown terminology}. Even though the research team was composed of people experienced in nursing culture, they were not always aware of the terms that were used by on-the-ground nurses.

\aptLtoX[graphic=no,type=html]{\begin{quote}
We search based on these weird, predefined keywords that we think relate to structural inequities. But we also have to guess in advance what language people might be using...We didn't think that people would delicately [post using the term] `people who experience structural inequities`...But we did try a lot of words we wouldn't use, you know, like addict, junkie...We were trying to use the system in a way that would get us further than those keywords alone. (P1)
\end{quote}}{\begin{italicquote}
We search based on these weird, predefined keywords that we think relate to structural inequities. But we also have to guess in advance what language people might be using...We didn't think that people would delicately [post using the term] `people who experience structural inequities`...But we did try a lot of words we wouldn't use, you know, like addict, junkie...We were trying to use the system in a way that would get us further than those keywords alone. (P1)
\end{italicquote}}

\textit{Teleoscope} helped to populate their keyword search document with search terms that they would not have predicted \textit{a priori} and helped them discover terms for phenomena of interest the researchers had not thought of.\\

\noindent
\textbf{\textit{Teleoscope} helped with search saturation}. The researchers reported (1) making groups from documents from their keyword searches; (2) piping the groups into the Ranking operation as controls; (3) determining which documents they had not yet read and (4) adding those documents to the groups. This helped uncover parts of the document space that they had not yet captured with a keyword search.

\aptLtoX[graphic=no,type=html]{\begin{quote}
Putting each one of our single groups like indigenous, vulnerable disabilities, [an RA] put them into the [Rank] and then basically went through to see like which ones we hadn't read...We were just trying to expand our data set and be exhaustive. (P1)\\
\end{quote}}{\begin{italicquote}
Putting each one of our single groups like indigenous, vulnerable disabilities, [an RA] put them into the [Rank] and then basically went through to see like which ones we hadn't read...We were just trying to expand our data set and be exhaustive. (P1)\\
\end{italicquote}}

We found that a process of crystallization~\cite{ellingson2009engaging} took place. Starting from a single document as a nucleation site, researchers started ``growing'' a more complex and interrelated document structure. Some documents were signposts (exceptions that proved a rule) which acted as inflection points. One researcher used a metaphor of being dropped off multiple times in a mountainous landscape, using certain documents as signposts along a pathway.

\noindent
\textbf{Working iteratively between keyword searches and ML functionality was important for exploring and structuring the research topic}. The researchers reported that they used the system to iterate and structure their ideas about their research topic.

\aptLtoX[graphic=no,type=html]{\begin{quote}
It was actually really helpful to start with keyword searches. And to be able to build out this groups structure, and then [Rank] from there. Whereas maybe if we'd had a more like drilled in topic, we could have just gone from there.(P1)
\end{quote}}{\begin{italicquote}
It was actually really helpful to start with keyword searches. And to be able to build out this groups structure, and then [Rank] from there. Whereas maybe if we'd had a more like drilled in topic, we could have just gone from there.(P1)
\end{italicquote}}

The researchers used \textit{Teleoscope} to develop increasingly sophisticated schemas that gradually superseded their original topical approach as their understandings of the target data grew:

\aptLtoX[graphic=no,type=html]{\begin{quote}
There are a bunch of different like origin categories we needed to go off of because of the way our topic is. We couldn't even [try to search for categorical terms such as] `Oh, this is about emergency departments...vulnerable [people] or inequities'...because when people are dragging on someone who uses drugs, who comes to the emergency department every week. That's not the words they're going to use. They're going to use super stigmatizing language, probably like what we've seen in a lot of cases and be like, `I'm so frustrated. This junkie comes into work all the time. He's just drug seeking. He's like plugging up a bed for everyone else who needs it.' (P1)\\
\end{quote}}{\begin{italicquote}
There are a bunch of different like origin categories we needed to go off of because of the way our topic is. We couldn't even [try to search for categorical terms such as] `Oh, this is about emergency departments...vulnerable [people] or inequities`...because when people are dragging on someone who uses drugs, who comes to the emergency department every week. That's not the words they're going to use. They're going to use super stigmatizing language, probably like what we've seen in a lot of cases and be like, `I'm so frustrated. This junkie comes into work all the time. He's just drug seeking. He's like plugging up a bed for everyone else who needs it.' (P1)\\
\end{italicquote}}

\noindent
\textbf{Projections operations can allay methodological concerns}. In the focus group, it was noted that using \textit{Teleoscope} might be too close to performing analysis rather than just curation, confounding methodological rigour (data curation and analysis stages are usually kept distinct so as not to predetermine results). However, the projection operation, which was added as a feature after the focus group, seemed to have potential to populate the interface with unexpected results (e.g., the signposts discussed in Section~\ref{sec:Formative Study}):

\aptLtoX[graphic=no,type=html]{\begin{quote}
One of the things that I had really worried about methodologically was that with \textit{Teleoscope}, I almost felt like you were kind of like deciding what your findings might be...I feel like [the projection operation] is really addressing some of that for me, because I feel like it's bringing you all these like adjacent topics to what you're looking for. And I feel like it really broadens your idea so much further in the potential data [since] you're exposed to so much more of the subreddit than you would be through just keyword searching. And I feel like that's really methodologically sound. (P1)
\end{quote}}{\begin{italicquote}
One of the things that I had really worried about methodologically was that with \textit{Teleoscope}, I almost felt like you were kind of like deciding what your findings might be...I feel like [the projection operation] is really addressing some of that for me, because I feel like it's bringing you all these like adjacent topics to what you're looking for. And I feel like it really broadens your idea so much further in the potential data [since] you're exposed to so much more of the subreddit than you would be through just keyword searching. And I feel like that's really methodologically sound. (P1)
\end{italicquote}}

The speed and ease of use of the Projection operation helped with researchers' sense of completion as well:

\aptLtoX[graphic=no,type=html]{\begin{quote}
When it comes to big data sets, it takes way too long to get a sense of what's going on. [The projection operation] is really nice, quick [and has] potential to expose that kind of stuff so that your findings aren't [close to the] single keyword that you put out...With qualitative research that's a really interesting and powerful thing to be able to do. (P1)\\
\end{quote}}{\begin{italicquote}
When it comes to big data sets, it takes way too long to get a sense of what's going on. [The projection operation] is really nice, quick [and has] potential to expose that kind of stuff so that your findings aren't [close to the] single keyword that you put out...With qualitative research that's a really interesting and powerful thing to be able to do. (P1)\\
\end{italicquote}}

\noindent
\textbf{Multiple workflows/workspaces allowed for reproducible exploration and collaboration.} The research team approached collaboration by exploring on their own in independent workspaces, discussing their results, and then collating the results into a single shared workspace. Outside collaboration tools were also used, such as email, Zoom, and Google Docs. The independent workspaces served as drafting areas, where individual researchers could explore many ideas without committing to the larger team's conception of the dataset, then came together with refined groups and keywords. After they collated their results, they performed further data exploration as a team on a single workspace by systematically using our ML operations to ensure completeness. Researchers were also able to compare their curation strategies and pathways with one another in order to surface different interpretive positions towards the data and examine how different curatorial choices led to resultant schemas.

Along with the above findings, researchers submitted a variety of feature requests and bug reports. Set operations were the last feature to be developed after some discussion and redesign of our backend graph processing system.

\subsubsection{Field Deployment: Conclusion}
In this evaluation, we performed a customized field deployment for a real qualitative research team and took ongoing bug reports and design feedback. Summarizing our findings in relation to our research questions:

\begin{itemize}[leftmargin=*]
    \item \textbf{RQ3 Result}. Researchers incorporated \textit{Teleoscope} into their research practice as a data curation tool, working between an external Google doc for keywords and the interface itself. \textit{Teleoscope} was used to explore parts of the document space where keywords would not be easy or obvious to find. 
    \item \textbf{RQ4 Result}. \textit{Teleoscope} helped researchers feel confident that a corpus was being more comprehensively and rigorously explored by providing both familiar and signpost example documents.
\end{itemize}

\subsection{Continued Public Release}
\textit{Teleoscope} continues to be maintained for use by the original research team, who partnered with their institution to evaluate potential market opportunities for \textit{Teleoscope} as a commercial system. We provide a brief account of our market explorations here to demonstrate that the core concept of \textit{thematic curation} uncovered in the \textit{Teleoscope} design process met the extensibility criterion of RtD \cite{zimmerman2007research} and users in domains other than the ones listed in the above evaluations envisioned being able to use the system to support their own data sense-making needs.

We provided demos of \textit{Teleoscope} and solicited feedback on whether users within a market research consultancy firm and a public policy think tank could envision using \textit{Teleoscope} to support their daily work with data. The authors of the system were also provided with some archetypal data sense-making tasks conducted by these organizations and used \textit{Teleoscope} to accomplish them. For market researchers, primary datasets consisted of large tabular Excel spreadsheets of market data that required an associated label to be assigned to them, in a process akin to sentiment analysis. When using \textit{Teleoscope}, the mental model of \textit{thematic curation} altered the way in which this labeling work was done. By iteratively refining \textit{thematic schemas}, labels could be assigned to a group of documents at once, rather than row-by-row processing. However, one challenge was to associate this kind of interpretive work back into the original tabular format to enable downstream business decision-making, so we implemented interfaces into \textit{Teleoscope} that allowed ingesting more archetypal data sources used by such organizations such as Excel spreadsheets and CSV files. Analysts from the public policy think tank had the research team use \textit{Teleoscope} on a different pool of social media data that came from X. 

\section{Discussion}
\label{sec:Discussion}
In this section, we discuss how \textit{Teleoscope's} approach to data curation helps support greater epistemological coherence between interpretivist research paradigms and working with large text corpora; how \textit{Teleoscope's} visualization approach affords qualitative approaches to data curation; and limitations and future directions for developing the core concepts of \textit{Teleoscope}.

\subsection{Supporting epistemologically coherent research with \textit{Teleoscope}}
Part of the motivation for developing \textit{Teleoscope} stemmed from a desire to provide an interpretivist approach to curating large text corpora from social media platforms like Reddit. Currently, curation and partitioning strategies in the literature \cite{low2020natural, slemon2021reddit, gauthier2022will} often employ ad-hoc strategies for curation prior to the main interpretive phase of their work in the analysis. In recent years, there have been concerns that interpretive HCI work is being increasingly evaluated and framed from a positivist position \cite{soden2024evaluating, bowman2023using}, resulting in inappropriate and incoherent critiques of interpretivist research questions. We observed a similar problem of epistemic incoherence stemming from the use of statistical approaches to data curation followed by interpretivist approaches to data analysis. This was exemplified even in our own design process in our move away from dashboard-style interfaces over the course of our first formative study.

Borrowing terminology from Kidder \& Fine \cite{kidder1987qualitative}, statistical sub-sampling approaches to data curation align more with what the authors refer to as "small-q" qualitative research: Qualitative research conducted within a \textit{quantitative} paradigm, as opposed to "Big-Q" qualitative research that embraces a "fully qualitative" approach. \textit{Teleoscope} aims to support qualitative methods within a qualitative paradigm, embracing researcher subjectivity as a fundamental part of knowledge construction \cite{bowman2023using}, or in \textit{Teleoscope's} case, knowledge curation. Without this, the combination of a "small-q" approach to data curation followed by a "Big-Q" approach to analysis might result in what Braun, Clarke, \& Hayfield describe as "confused q" qualitative research \cite{braun2021one}, "research that seems to unknowingly, unreflexively and incoherently combine elements of qualitative positivism with the values and assumptions of a qualitative paradigm". \textit{Teleoscope's} aims to support a "Big-Q" approach to data curation, enabling a consistent and coherent epistemological foundation for interpretivist research projects that wish to make use of large text corpora as their site of inquiry.

However, it would be over-claiming to state that \textit{Teleoscope} ensures full interpretivist rigour in the traditional sense. After all, this would necessitate manually examining the full dataset, the original problem \textit{Teleoscope} was trying to probe a solution to. Instead, we see the \textit{Teleoscope} design process as a response to the call by Baumer to develop systems that "provide a scaffolding for
human interpretation" \cite{Baumer2017}, which is exactly what \textit{thematic schemas} are intended to do, allowing for model outputs that "may actually be more valuable if they provide some uncertainty and interpretive flexibility" \cite{Baumer2017}. \textit{Teleoscope} is intentionally designed not to over-step into the qualitative analysis process (see Figure\ref{fig:teleoscope-process}), allowing qualitative researchers full interpretive control over their actual analysis procedure. It is true that \textit{Teleoscope} still requires researchers to imbue the "impressionistic" process \cite{ramsay2003reconceiving} of interpretivist data curation with computational notions of similarity (via vector similarity). However, the core human-in-the-loop interaction of iteratively refining abstract mental models about data and reifying them into \textit{thematic schemas} remains fundamentally committed to interpretivist approaches of working with data.

\subsection{Deductive vs. Inductive Approaches to Corpus Exploration}
Over the course of working on \textit{Teleoscope}, the research team realized that supporting an interpretivist approach to data curation required a fundamental shift in the kind of visualization support provided to the researcher. Prior systems for large-scale data exploration, such as Scholastic \cite{Hong2022}, Serendip ~\cite{Alexander2014}, and TopicSifter ~\cite{kim2019topicsifter},  focus on "top-down" or deductive approaches to developing schemas about the corpus. They typically rely on visualization at the "corpus-level", presenting broad visual overviews of the dataset before allowing user's to drill down or converge to specific documents or topics. In contrast, \textit{Teleoscope} takes a "bottom-up" or inductive approach to assisting with schema development, which influenced the sparse, canvas style interface we deployed to participants. By visualizing the corpus at the "document-level", researchers are able to curate at a level close to the data itself and explore in a divergent manner, rather than having to make interpretive jumps by reasoning about the specific documents that might be contained within a large collection, as is typical with the more deductive/convergent approach.

Another advantage of a more inductive approach to corpus visualization and curation is the interpretive flexibility and freedom it affords to users. Instead of working with overviews of the full corpus, which might be unwieldy, or machine-generated topic clusters, which might not have any relation to the actual interpretive categories of interest for a researcher \cite{endert2012semantics}, the inductive approach allows researchers to iteratively explore data without foreclosing any particular curative pathways. This ability to support immersion within the data \cite{braun2012thematic} can lead to creative and serendipitous encounters with the data, an important aspect identified by Jiang et. al \cite{jiang_supporting_serendipity} when introducing AI-powered assistance into the qualitative research life-cycle. This again helps \textit{Teleoscope} support an "end-to-end" interpretivist approach to qualitative research from the curation to the analysis stage.

\subsection{Benefits and Limitations of \textit{Teleoscope's} usage metaphor and workflow}

We found that the core benefit of Teleoscope was that it provided researchers with confidence that they had explored a large document space with confidence, and apply the values of thematic analysis to curation. By using a visible trace of thematic schemas, they could re-inspect and share their thinking processes, ensuring that their curated documents provided saturation and that their curation was rigorous. Through working directly with qualitative researchers in an extended design process, we discovered methodological processes unique to large-scale semantic search for thematic curation, such as signposting and crystallization to describe boundaries and inflection points between document sets. 

Like an affinity diagram at scale, Teleoscope facilitated trust through the process of arranging documents. Rather than "offloading the work" to an LLM, Teleoscope provided the virtual material for researchers to manipulate, structure and capture their thought processes at a scale that would not be possible by hand. By allowing researchers to impose their own meaning on document similarity searches, Teleoscope allowed interpretivist approaches to make headway into statistical—often hidden—curation methodologies at scale.

Through our studies and design work, we settled on a process-focused graph-based workflow metaphor rather than a dashboard metaphor (\textbf{DG2}). This was due to our focus on the process of arranging and connecting documents as being the most important interaction focus (\textbf{DG1}, \textbf{DG3}); it also increased visibility of system state and allowed for direct inspection of computational elements. However, one limitation of a workflow metaphor is that it does not visually model the unknown document space very well. In part, this helps manage cognitive overload since users are only shown what they have explicitly searched for themselves (as we argue above). However, some dashboard elements, such as summary statistics of corpus exploration relative to selected groups, could provide some assistance to users if they are presented on-demand in response to curatorial needs.
Due to our users' use of an external keyword Google doc in the field deployment, an open question is how and whether to incorporate keywords back into \textit{Teleoscope} as a primary interaction element (\textbf{CS2.RQ1}). Perhaps with more experience with \textit{Teleoscope}, set operations could be used to manage keywords more directly. 

In addition, the implementation of \textit{Teleoscope} explored in this paper does not make use of general-purpose LLMs (such as OpenAI's GPT or Anthropic's Claude family of models) to our computational features such as Search, Rank, and Projection. That being said, the \textit{Teleoscope} code base is open-source \cite{teleoscope2024githubpostreview} and supports the ability to "plug and play" other models. Examining whether the incorporation of lightweight, general-purpose LLMs improves semantic filtering quality while continuing to support our design goals of continuous interactivity and interpretive immersion is a logical next step for future work. Another surprising limitation we uncovered through our deployments of \textit{Teleoscope} (particularly the second, extended formative study) was how steep the learning curve of the interaction metaphors seemed to be for qualitative researchers, a notion that has been empirically supported by research on qualitative research processes \cite{jiang_supporting_serendipity}. Qualitative researchers are somewhat resistant to \textit{learning} new computational tooling. We conjecture this could possibly be due to a lack of proper incentive structures (reporting using a relatively unproven and new method can be challenging for a researcher to justify, particular researcher's early in their career). With the advent of LLMs, we might also consider providing some interactive ways of on-boarding new users into the system, as some researchers have proposed for qualitative analysis tasks \cite{hayes2025conversing}.

Lastly, although the paper used Reddit as the primary example of a qualitative site containing large text corpora, the authors recognize that significant changes to its API since 2023 as a result of the advent of ChatGPT have severely limited researcher access to site data. Despite this restriction, \textit{Teleoscope} is not designed specifically for use with Reddit data, and the research team has since implemented the ability for users to upload their own data sources and corpora in response to changes in Reddit's API. In our opinion, a large part of \textit{Teleoscope's} success is also demonstrated by its on-going public release and use by the research team that we recruited for our field deployment. We are continuing to recruit more research teams and hope to see \textit{Teleoscope} develop into a more mature product as researchers adopt it.

\section{Conclusion}
\label{sec:Conclusion}
We present \textit{Teleoscope}, a web-based system that supports interpretivist approaches to curation of large text corpora for downstream analysis, a process we call thematic curation. We developed the system in response to the needs of qualitative researchers to explore large corpora in ways that were coherent with interpretivist paradigms of HCI research.
\textit{Teleoscope} provides NLP-based operations on a visualization interface that supports many of the interaction patterns familiar to qualitative researchers. \textit{Teleoscope} facilitates the creation of thematic schemas from large text corpora in a way that allows researchers to retrace, share, and recompute their sense-making process to other researchers.
We reported on the design process, engineering, evaluation, and deployment of our system. Our public deployment of \textit{Teleoscope} is ongoing~\cite{teleoscope2024postreview} and we plan to continue improving \textit{Teleoscope} and to maintain it for use by the broad community of qualitative researchers.

\begin{acks}
We would like to thank our study participants, without whom this work would not have been possible. A heartful thanks also to the numerous individuals who worked on the \textit{Teleoscope} project over the years, including Aanandi Sidharth, Qiyu Zhou, Kenny Averna, Dhruv Khanna, Sol Lee, Crystal Lee, Prayus Shrestha, Florentina Simlinger, and Vita Chan.

We acknowledge the Natural Sciences and Engineering Research Council of Canada (NSERC), [grant number RGPIN-2020-05203, CGS-D, CGS-M] for their support. This work was also supported by the DFP Project Stimulus Grant, the British Columbia Graduate Scholarship, and Huawei Research.
\end{acks}

\bibliographystyle{ACM-Reference-Format}
\bibliography{bibliography}


\end{document}